\documentstyle[epsf,wrapft]{ptptex}
\newcommand{\bra}[1]{\left\langle #1 \right|}         
\newcommand{\ket}[1]{\left| #1 \right\rangle}         

\newcommand{\de}[2]{{d #1 \over d #2}}                
\newcommand{\pa}[2]{{\partial #1 \over \partial #2}}  
\newcommand{\papa}[3]{{\partial^2 #1 \over \partial #2 \partial #3}} 
						      
\newcommand{\fd}[2]{{\delta #1 \over \delta #2}}      
\newcommand{\fdfd}[3]{{\delta ^2 #1 \over \delta #2 \delta #3}} 
                                                     
\newcommand{\ska}[1]{\left( #1 \right) }             
\newcommand{\mka}[1]{\left\{ #1 \right\} }           
\newcommand{\bka}[1]{\left[ #1 \right] }             
\newcommand{\lsim}{\raisebox{0.2ex}{$\ < \kern -1.05em%
        \raisebox{-1.1ex}{$\sim$}\ $}}  % =<
\newcommand{\gsim}{\raisebox{0.2ex}{$\ > \kern -1.05em%
        \raisebox{-1.1ex}{$\sim$}\ $}}  % >=

\notypesetlogo  %comment in if to eliminate PTPTeX logo
\title{%        %You can use \\ for explicit line-break
Non-Perturbative Renormalization Group Analysis \\
in Quantum Mechanics
}
\author{%       %Use \sc for the family name
Ken-Ichi {\sc Aoki},\footnote{E-mail: aoki@hep.s.kanazawa-u.ac.jp}
Atsushi {\sc Horikoshi},\footnote{E-mail: horikosi@hep.s.kanazawa-u.ac.jp}
Masaki {\sc Taniguchi}\footnote{E-mail: taniguti@snc.sony.co.jp} \\and
Haruhiko {\sc Terao}\footnote{E-mail: terao@hep.s.kanazawa-u.ac.jp} 
}

\inst{%         %Affiliation, neglected when [addenda] or [errata]
Institute for Theoretical Physics, Kanazawa University,\\
Kanazawa 920-1192, Japan
}

\abst{%       %this abstract is neglected when [addenda] or [errata]
We analyze quantum mechanical systems 
using the non-perturbative renormalization group (NPRG).
The NPRG method enables us to calculate quantum corrections 
systematically and is very effective for studying non-perturbative dynamics.
We start with anharmonic oscillators and proceed to 
asymmetric double well potentials,
supersymmetric quantum mechanics and many particle systems.
}

\begin{document}

\maketitle

\section{Introduction}
The non-perturbative renormalization group (NPRG) has been formulated 
through analyses of critical phenomena\cite{wk}
and applied to non-perturbative studies of
statistical mechanics and quantum field theories.
It has been established as a powerful tool for analyses of 
non-perturbative dynamics in systems of many (infinite) degrees of
freedom,
because it allows for the evaluation of fluctuations without recourse 
to perturbation series.
Several types of non-perturbative ({\it exact}) renormalization group
equations have been derived by integration with scale decomposition 
and have been applied to 
various systems.\cite{aoki1,mo,jw,erg1,aoki2,bb,erg2,btw}
In this article, we apply the NPRG method to 
quantum mechanical systems, that is, systems of 
finitely many degrees of freedom\cite{aoki1,aoki2,ahtt}  
to analyze their non-perturbative dynamics.
\par
Generally, there are two types of non-perturbative quantities. One
corresponds to the summation of all orders of a perturbative series,
which could be related to Borel resummation.\cite{zj} 
The other is an essential singularity with respect to a coupling
constant $\lambda_0$, which has a structure
like $e^{-\frac{1}{\lambda_0}}$.\cite{co,kl}
We are not able to expand such a singular 
contribution around $\lambda_0=0$. A singularity of this type 
appears in the case of quantum tunneling. For example, in a
symmetric double well system, there are two degenerate energy levels
at each minimum, which are mixed through tunneling 
to generate an energy gap $\Delta E \sim e^{-\frac{1}{\lambda_0}}$. 
The exponential factor is known to result from the free energy of topological
configurations, i.e., instantons. 
\par
In this article, we first summarize how to analyze quantum mechanical
systems using the concept of NPRG and check to what extent NPRG 
can be used to  
evaluate non-perturbative effects quantitatively. 
The NPRG equation we employ here is a local potential approximated 
Wegner-Houghton (LPA W-H) equation,\cite{wh,ap} which we use to 
analyze quantum anharmonic oscillators and asymmetric double well
systems.
In contrast to the symmetric double well system, the standard instanton
method does not work for an asymmetric potential, and the much more 
sophisticated method of the valley instanton 
has been developed for their treatment.\cite{ao1,ao2}
The NPRG method is found to work for asymmetric potentials as 
well as for symmetric potentials, because NPRG does not rely on parity 
symmetry.
\par
We proceed to analyses of more complicated systems, 
supersymmetric quantum
mechanics (SUSY QM) and many particle systems.
SUSY QM is a toy model for dynamical SUSY 
breaking.\cite{wi1,sv}
Although, in general, there is no spontaneous symmetry breaking in 
systems with finitely many degrees of freedom, some extraordinary 
symmetries, such as SUSY, can be broken even in quantum mechanics.
SUSY breaking is a highly non-perturbative phenomenon 
because of the non-renormalization theorem, 
and we will see that NPRG should be applicable for non-perturbative 
SUSY breaking.
\par
In addition, analyses of quantum many particle 
systems have become very important 
with recent developments in nano-technology.
Solving the problem of how the quantum coherence of 
a variable of a target system 
is affected by other variables (the environment) is quite important.
For example, it is necessary for realization of {\it qubit} 
for quantum computers.
However, standard methods that are well suited 
for treating systems of one degree of freedom, 
the Schr\"odinger equation, instanton, etc.,
do not work well in such complicated systems.   
We believe that NPRG is versatile enough to analyze such systems.
As a first step, we analyze quantum tunneling phenomena in 
two particle quantum systems.  
\par
\section{Non-perturbative renormalization group}
In this section, we briefly summarize the formulation of NPRG with
$D$-dimensional real scalar field theory.   
\subsection{Scale decomposition}
In the NPRG method, the theory is defined by 
the Wilsonian effective action $S_{\Lambda}[\phi]$.
This is an effective theory with 
an ultraviolet energy cutoff $\Lambda$:
 \begin{eqnarray}
 Z=\int {\cal D}\phi~e^{-S_\Lambda[\phi]}.\label{(1)}
 \end{eqnarray}
We decompose the path integration variable $\phi (p)$ into two parts 
with respect to the momentum scale $p$ as
 \begin{eqnarray*}
   \phi (p)=\left\{
    \begin{array}{ll}
     \phi_{<} (p) &~~~~~~~~~~~~\!0\leq|p|< \Lambda -{\mit\Delta}\Lambda
       ~~:{\rm lower~modes}, 
        \\
     \phi_s ~(p) &~~~\Lambda-{\mit\Delta}\Lambda\leq|p|\leq 
       \Lambda ~~~~~~~~
       ~~:~{\rm shell~modes},
    \end{array}
    \right.
 \end{eqnarray*}
and transform the partition function $Z$ as follows:
 \begin{eqnarray}
 Z&=&\int {\cal D}\phi_{<}~{\cal
 D}\phi_{s}~e^{-S_{\Lambda}[\phi_{<}+\phi_{s}]}, \nonumber\\
&=&\int {\cal D}\phi_{<}~e^{-S_{\Lambda}[\phi_{<}]}
\int {\cal D}\phi_{s}~e^{-S_{\Lambda}[\phi_{s}]}~
e^{-S_{\Lambda}^{\rm ~\!int}[~\!\phi_{<},~\phi_{s}~\!]},\nonumber\\
&=&\int {\cal D}\phi_{<}~e^{-S_{\Lambda}[\phi_{<}]}
\left<~e^{-S_{\Lambda}^{\rm ~\!int}[~\!\phi_{<},~\phi_{s}~\!]}\right>
_{\phi_{s}},\nonumber\\
&=&\int {\cal D}\phi_{<}~e^{-S_{\Lambda}[\phi_{<}]}
~e^{- {\mit\Delta}S_{\Lambda}[\phi_{<}]},\nonumber\\
&=&\int {\cal D}\phi_{<}~e^{-S_{\Lambda-{\mit\Delta}\Lambda}[\phi_{<}]},\label{(2)}
 \end{eqnarray}
where  
 \begin{eqnarray}
{\mit\Delta}S_{\Lambda}[\phi_{<}]
\equiv-\log\left<~e^{-S_{\Lambda}^{\rm ~\!int}
[~\!\phi_{<},~\phi_{s}~\!]}\right>_{\phi_{s}}
\equiv\int {\cal D}\phi_{s}~e^{-S_{\Lambda}[\phi_{s}]}~
e^{-S_{\Lambda}^{\rm ~\!int}[~\!\phi_{<},~\phi_{s}~\!]}.\label{(3)}
 \end{eqnarray}
We understand the shell mode path integral 
$\left<\cdot\cdot\cdot\right>_{\phi_{s}}$
as the renormalization transformation.
If we evaluate it by perturbative expansion with respect to 
coupling constants, we obtain the 
so-called perturbative renormalization group 
equations.\cite{ps} Of course such equations are valid only in the weak
coupling limit.
Instead, we take the limit  
${\mit\Delta}\Lambda\to0$ to define NPRG equation,
which is the fundamental procedure.\cite{wh}
\subsection{Derivation of the NPRG equation}
Taking the limit ${\mit\Delta}\Lambda\to0$, we can express 
the renormalization transformation as a differential equation,  
 \begin{eqnarray}
  \frac{\partial S_{\Lambda}}{\partial
  \Lambda}=\lim_{{\mit\Delta}\Lambda \to 0}\frac{S_{\Lambda}-
S_{\Lambda-{\mit\Delta}\Lambda}}{{\mit\Delta}\Lambda}
=\lim_{{\mit\Delta}\Lambda \to 0}\frac{1}{{\mit\Delta}\Lambda}
~{\rm log}~\left<~e^{-S_{\Lambda}^
{\rm ~\!int}[~\!\phi_{<},~\phi_{s}~\!]}\right>
_{\phi_{s}}.\label{(4)}
 \end{eqnarray}
Then, we can expand $S_{\Lambda}[\phi]$ as power series in $\phi _s$:
 \begin{eqnarray}
  S_{\Lambda} \left[ \phi \right] &=& S_{\Lambda} \left[ \phi_{<} 
   \right] 
  +\int_{\rm shell} \left.\fd{S_{\Lambda}}
{\phi \left( p \right)}\right| 
   _{\phi_s=0} \!\!\!\!\!\!\cdot \phi_s \left( p \right) \nonumber\\
  && +~{1 \over 2}\int\int_{\rm shell}
    \phi_s (p)\cdot\left.\fdfd{S_{\Lambda}}{\phi (p)}{\phi (q)}
   \right| _{\phi _s =0} \!\!\!\!\!\!\cdot \phi_s(q)~
+~O({\mit\Delta}\Lambda ^2 ).\label{(5)}
 \end{eqnarray}
Since an $O({\mit\Delta}\Lambda)$ calculation is required for 
the evaluation of the derivative (\ref{(4)}),
the shell mode path integral $\left<\cdot\cdot\cdot\right>_{\phi_{s}}$
can be evaluated {\it exactly} using a Gaussian integration.
Then, the fundamental differential equation is derived as
\begin{eqnarray}
 \Lambda \frac{\partial S_{\Lambda}}{\partial\Lambda}
=~\frac{\Lambda}{2}\int_{\rm shell}
\left\{
-~{\rm log}\left(\left.\fdfd{S_{\Lambda}}{\phi_{p}}{\phi_{-p}}
   \right|\right)
+\left.\fd{S_{\Lambda}}{\phi_{p}}\right|
\left(\left.\fdfd{S_{\Lambda}}{\phi_{p}}{\phi_{-p}}
   \right|\right)^{-1}\!\!\!
\left.\fd{S_{\Lambda}}{\phi_{-p}}\right|
~\right\}.\label{(6)}
 \end{eqnarray}
This is known as the Wegner-Houghton equation.\cite{wh}
It represents exactly the cutoff $\Lambda$ dependence of 
the Wilsonian effective action $S_{\Lambda}$.
Its right-hand side is generally called a {\it $\beta$ functional}.
\subsection{Approximations}
Although the Wegner-Houghton equation is exact, we cannot
solve it without some approximation in practice.
In this article we employ the local potential approximation (LPA), 
which means that we ignore corrections to derivative interactions.
It can be considered the leading order of the 
derivative expansion of $S_{\Lambda}$, 
\begin{eqnarray}
S_{\Lambda}[\phi]=\int d^D x ~\left\{~V_{\Lambda}[\phi] 
+\frac{1}{2}K_{\Lambda}[\phi]\!~
\partial_{\mu}\phi \partial_{\mu}\phi
+\cdot\cdot\cdot\right\}.\label{(7)}
 \end{eqnarray} 
To make this approximation,   
we substitute the zero mode for the 
lower mode $\phi_{<}(p)$:
\begin{eqnarray}
\phi_{<}(p)\to\varphi ~(2\pi)^D\delta^D(p).\label{(8)}
 \end{eqnarray}
The local potential approximated Wegner-Houghton (LPA W-H) equation 
is then obtained as follows: 
 \begin{eqnarray}
 \Lambda\pa{V_{\Lambda}}{\Lambda}&=&
-~\!\frac{A_{D}}{2}~\!\Lambda^{D}
  ~\!\log\ska{1+\frac{1}{\Lambda ^2}\frac{\partial ^2
  V_{\Lambda}}{\partial \varphi ^2} },\\ \label{(9)}
A_D &\equiv&\frac{\int d\Omega _{D}}{(2\pi)^D}.\label{(9a)}
 \end{eqnarray}
This is a two-dimensional 
partial differential equation for 
$V_{\Lambda}(\varphi).$\cite{ap} 
Its right-hand side is called a {\it $\beta$ function.}
We solve it mainly using numerical methods.
\par
To obtain an intuitive understanding of this equation, 
we proceed to further approximation, the operator expansion.
We expand $V_{\rm eff}$ as power series in $\varphi$: 
\begin{eqnarray}
 V_{\Lambda} \left(\varphi\right) &=&
 \sum_{n=0}^N\frac{a_n(\Lambda)}{n!}\varphi ^n.\label{(10)}
\end{eqnarray}
The partial differential equation is then reduced 
to a set of ordinary differential
equations for the coupling constants $\{{a}_n(\Lambda)\}$: 
\begin{eqnarray}
\Lambda\de{{a}_0}{\Lambda}&=&-\frac{A_D}{2}~\!\Lambda^{D}~\!
\log\left(\frac{\Lambda^2+{a}_2}
{\Lambda^2}\right),\label{(11)}\\
  \Lambda\de{{a}_1}{\Lambda}&=&-\frac{A_D}{2}~\!\Lambda^{D}~\!
\left[{a}_3 \over
 \Lambda^2+{a}_2\right],  \label{(12)}\\
  \Lambda\de{{a}_2}{\Lambda}&=&-\frac{A_D}{2}~\!\Lambda^{D}~\!
\bka{{{a}_4 \over \Lambda^2 +{a}_2}-
    {{a}_3^2 \over \ska{\Lambda^2+{a}_2}^2}}, \label{(13)}\\
  \Lambda\de{{a}_3}{\Lambda}&=&-\frac{A_D}{2}~\!\Lambda^{D}~\!
\bka{{{a}_5 \over \Lambda^2+{a}_2}-{3{a}_4{a}_3
    \over \ska{\Lambda^2+{a}_2}^2}
+{2{a}_3^3 \over \ska{\Lambda^2+{a}_2}^3}}, \label{(14)}\\
  \Lambda\de{{a}_4}{\Lambda}&=&
-\frac{A_D}{2}~\!\Lambda^{D}~\!
\left[{{a}_6 \over \Lambda^2+{a}_2}-{4{a}_5{a}_3
    \over \ska{\Lambda^2+{a}_2}^2}
-{3{a}_4^2 \over \ska{\Lambda^2+{a}_2}^2}\right.\nonumber\\
&&~~~~~~~~~~~~~~~~~~~~~~~~~~~~~~~~~~~~~~~~~~~
\left.+\frac{12{a}_4{a}_3^2}{\ska{\Lambda^2+{a}_2}^3} 
-\frac{6{a}_3^4}{\ska{\Lambda^2+{a}_2}^4}
\right]
, \label{(15)}\\
 && \vspace{3cm} \vdots
~~~~~~~~~~~~~~~~~~~~~~~~~~~~~~~~~~~~~~~~~~~~~~~~~
~~~~~~~~~~~~~~~~~~~~~~~~~~~~~~~. \nonumber
 \end{eqnarray}
If the results of these equations converge as the 
order of the truncation, $N$,
becomes large, we regard them as solutions 
of the LPA W-H equation.\cite{ka} 
This expansion allows us to treat differential equations more easily
and to understand the origin and structure of quantum corrections
physically.
For example, a correction to odd $n$ couplings cannot be generated
from even $n$ couplings only.
This implies that if we choose an initial potential $V_0(\varphi)$ as
$Z_2(\varphi\leftrightarrow-\varphi)$ symmetric, 
the solutions of the RG equations flow in $Z_2$ symmetric subspace; 
that is, the NPRG equation does not break the global symmetry of the system.
However, we should note that in some extreme cases, the 
operator expansion is not good, and leads to pathological behavior,
as seen in \S 4.  
\par 
The constant part of $V_{\Lambda}$, $a_0$, is given by the vacuum
bubble diagrams\footnote{
For other types of NPRG equations, such as 
the Legendre flow equation, which is derived by means of 
a cutoff function,\cite{we2} we cannot evaluate the vacuum
bubble diagrams properly without the prescription of
subtracting the contribution of the cutoff function from the 
constant part of $V_{\Lambda}$.} 
and is usually ignored. However, we retain it here,
because it plays a crucial role in supersymmetric theories.  
\par
\section{NPRG analysis of quantum mechanical systems}
Making use of the LPA W-H equation, we analyze 
systems in quantum mechanics,
which is $D=1$ real scalar theory 
with a single dynamical variable, $x(\tau)$.\cite{ahtt}

\subsection{Physical quantities}
The LPA W-H equation for quantum mechanics is given as follows:
 \begin{eqnarray}
 \Lambda\frac{\partial V_{\Lambda}}{\partial \Lambda}&=&
-~\!\frac{1}{2\pi}~\!\Lambda
  ~\!\log\ska{1+
\frac{1}{\Lambda ^2}\frac{\partial ^2
  V_{\Lambda}}{\partial x ^2} }.\label{(16)}
 \end{eqnarray}
We solve it by lowering $\Lambda$ 
from the initial cutoff $\Lambda _0$, 
where the initial potential $V_{\Lambda_0}$ is given 
by the potential term $V_0(x)$ in the original action,
\begin{eqnarray}
S[x]=\int d \tau 
~\left\{~\frac{1}{2}\!~\dot{x}^2+V_0(x) 
\right\}.\label{(17)}
 \end{eqnarray} 
In the infrared limit $\Lambda\to 0$, we obtain the effective potential 
$V_{\rm eff}(x)=\lim_{\Lambda\to 0}V_{\Lambda}(x)$, from which 
physical quantities are evaluated.
\par
First, the expectation value of $x$ 
in the ground state $\ket{\Omega}$,
 \begin{eqnarray}
\langle x\rangle\equiv \bra{\Omega}\hat{x}\ket{\Omega},\label{(18)}
 \end{eqnarray}
is determined by the stationarity condition,
 \begin{eqnarray}
\left.\frac{d V_{\rm eff}}{d x}\right|_{x=\langle x\rangle}=0.\label{(19)}
 \end{eqnarray}
The ground state energy of the quantum system is given by
 \begin{eqnarray}
E_0=\bra{\Omega}\hat{H}\ket{\Omega}=
V_{\rm eff}(x=\langle x\rangle).\label{(20)}
 \end{eqnarray}
Also, we obtain 
the energy gap of the system 
through the following expressions of 
the two-point correlation function:
 \begin{eqnarray}   
\bra{\Omega}{T}\hat{x}(\tau)\hat{x}(0)\ket{\Omega}
&=&
\int
\frac{dE}{2\pi}e^{iE\tau}\sum_{n=1}^{\infty}\frac{D_n}{E^2+(E_n-E_0)^2}
\stackrel{\tau\to\infty}{\propto } e^{-(E_1-E_0)\tau}, \label{(21)}\\
\bra{\Omega}{T}\hat{x}(\tau)\hat{x}(0)\ket{\Omega}
&\stackrel{LPA}{=}&
\int \frac{dE}{2\pi}e^{iE\tau}\frac{1}{E^2+m^2_{\rm eff}}
=\frac{1}{2m_{\rm eff}}
e^{-m_{\rm eff}\tau},\label{(22)}
 \end{eqnarray}
where $D_{n}\equiv 2|C_n|^{2}(E_n-E_0),~C_{n}\equiv
\bra{n}\hat{x}(0)\ket{\Omega},~\sum_{n}D_n=1$ and 
$m_{\rm eff}^2$ is the curvature 
of the effective potential at the minimum. 
Comparing the damping factors of (\ref{(21)}) and (\ref{(22)}) 
in the $\tau\to\infty$ region,
the energy gap $\Delta E=E_1-E_0$ is obtained as follows:\footnote{
For further details with regard to this relation, see Appendix A.
}
 \begin{eqnarray}   
\Delta E =m_{\rm eff}=\sqrt{\left.\frac{\partial^2V_{\rm eff}}
{\partial x^2}\right|}_{x=\langle x\rangle}.\label{(23)}
 \end{eqnarray} 
\par
Generally, the NPRG method does not evaluate the wave functions of the
target system directly.
Instead, it yields information regarding Green functions.
For example, the $n$-th moment of $\hat{x}$ corresponds to 
information concerning the ground state wave function $\psi _0(x)$:
 \begin{eqnarray}
 M_n &=&
\langle \Omega |~\hat{x}^n| \Omega \rangle
=\int \!dx ~x^n \left| \psi _0(x)\right|^2.\label{(24)}
 \end{eqnarray}
We now give some examples of calculations of $M_n$ 
in a $Z_2$-symmetric (namely, $\langle x\rangle=0$) system.
The two-point function $M_2$ is calculated as
 \begin{equation}
  M_2=\langle \Omega |~\hat{x}^2| \Omega \rangle _c
\stackrel{LPA}{=}\int \frac{dE}{2\pi}\frac{1}{E^2+m^2_{\rm eff}}
  =\frac{1}{2m_{\rm eff}},\label{(25)}
 \end{equation}
where the subscript $c$ denotes a 
connected function.
In a similar way, the four-point function $M_4$,  
 \begin{eqnarray}
M_4=\langle \Omega |~\hat{x}^4| \Omega \rangle _c 
+3 M_{2}^{~2},\label{(26)}
 \end{eqnarray}
is calculated by means of 
the LPA four-point coupling $~\lambda_{\rm eff}
\equiv\frac{\partial^{4}V_{\rm eff}}{\partial x^{4}}$:
 \begin{eqnarray}
\langle \Omega |~\hat{x}^4| \Omega \rangle _c
&\stackrel{LPA}{=}&
-4!~\lambda_{\rm eff}\int\frac{dE_1dE_2dE_3}{(2\pi)^3}
\nonumber \\
&&\times   ~{1 \over E_1^2\!+m^2_{\rm eff}}
   ~{1 \over \ska{E_2-\!E_1}^2\!+m^2_{\rm eff}}
   ~{1 \over \ska{E_3-\!E_2}^2\!+m^2_{\rm eff}}
   ~{1 \over E_3^2\!+m^2_{\rm eff}},
\nonumber \\
&=&-{3\lambda_{\rm eff} \over 4m^5_{\rm eff}}.\label{(27)}
 \end{eqnarray}
In this way, we are able to connect the effective couplings
obtained from the NPRG with $M_n$ obtained from  
the ground state wave function.
All of this information concerning the ground state 
taken together provides 
full information regarding the quantum system, 
including excited states.
\subsection{Example: harmonic oscillator}
To illustrate the characteristics of the NPRG
analysis, we now consider the harmonic oscillator. 
The initial potential is chosen as $V_0(x)=\frac{1}{2}m^2 x^2$
at the initial cutoff scale $\Lambda_0$.
In this case, we can solve the LPA W-H equation analytically, and we obtain
\begin{eqnarray}
a_0 (\Lambda)&=&a_0(\Lambda_0)+
 \frac{\sqrt{a_2(\Lambda_0)}}{2\pi}
\left[\hat{p}\log \frac{1+\hat{p}^2}{\hat{p}^2}+2\tan^{-1}\hat{p}
 \right]^{\hat{p}
=\frac{\Lambda_0}{\sqrt{a_2(\Lambda_0)}}}_{\hat{p}
=\frac{\Lambda}{\sqrt{a_2(\Lambda_0)}}} , \label{(28)}\\
a_2(\Lambda)&=&a_2(\Lambda_0).\label{(29)}
\end{eqnarray}
Since the initial conditions are 
$(a_0(\Lambda_0),a_2(\Lambda_0))=(0,m^2)$,
if we take the simultaneous limit $\Lambda_0\to \infty$, $\Lambda\to 0$, 
$(a_0(\Lambda),a_2(\Lambda))=(\frac{m}{2},m^2)$ 
is obtained.
Although $a_2 (\Lambda)$ is free from quantum corrections 
and does not run, $a_0$ runs and  
produces a zero-point energy $\frac{m}{2}$. 
\par
\begin{wrapfigure}{r}{6.6cm}
\epsfysize=60mm
\epsfxsize=69mm
\leavevmode
\epsfbox{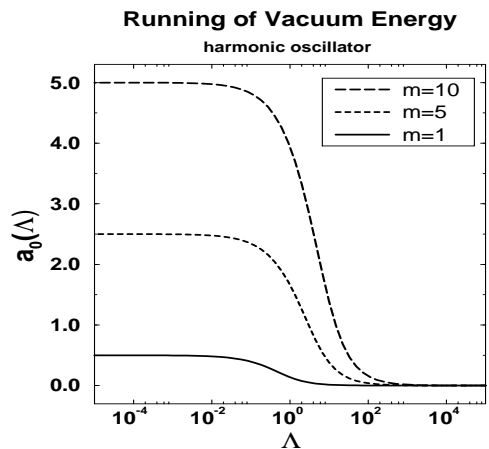}
\vspace{-6mm}
\caption{Running of $a_0$.}
 \label{fig:sirun}
\end{wrapfigure}
\par
Figure.\ref{fig:sirun} plots 
the actual running of $a_0$ and shows that it 
is limited to a finite energy region
that depends on the mass scale, $m$. 
Ultraviolet finiteness is a typical feature 
of quantum mechanical systems, and it implies that
the theory is finite, even in the $\Lambda_0\to \infty $ limit. 
Contrastingly, 
the infrared finiteness in Figure.\ref{fig:sirun} 
is related to the decoupling property 
that a heavy particle cannot propagate 
in the low energy region.
Such ultraviolet finiteness and infrared finiteness enable us to 
obtain physical quantities even through numerical calculation 
within a finite energy scale region.
\section{Analysis of anharmonic oscillators and double well systems}
\subsection{Symmetric single-well potential}
\begin{figure}[htb]
\hspace{0mm}
 \parbox{65mm}{
 \epsfxsize=65mm 
 \epsfysize=65mm
  \leavevmode
\epsfbox{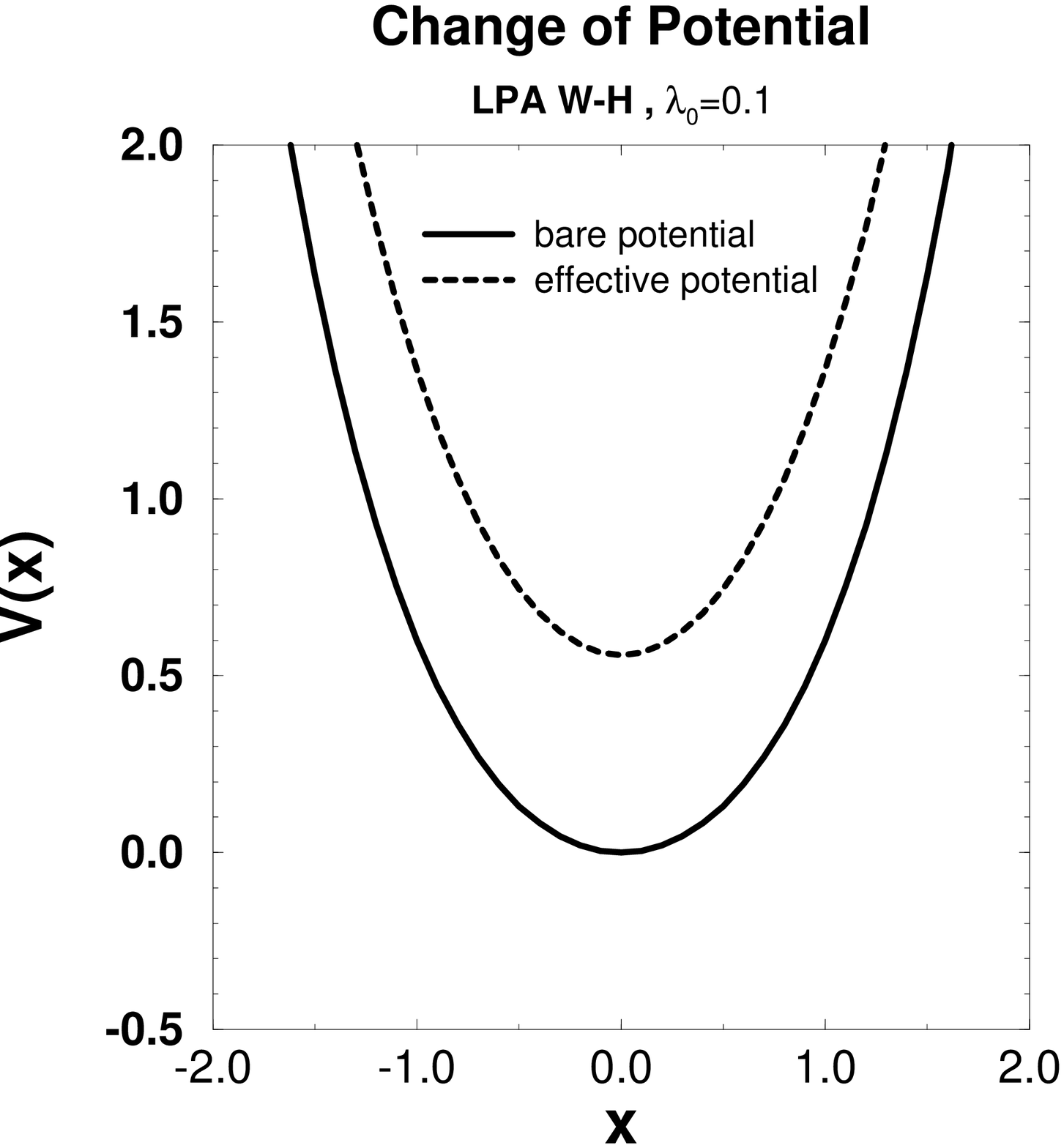}
\vspace{-3mm}
 \caption{Potential flow.}
 \label{fig:spote}
 }
\hspace{0mm} 
\parbox{65mm}{
 \epsfxsize=65mm 
 \epsfysize=65mm
 \leavevmode
\epsfbox{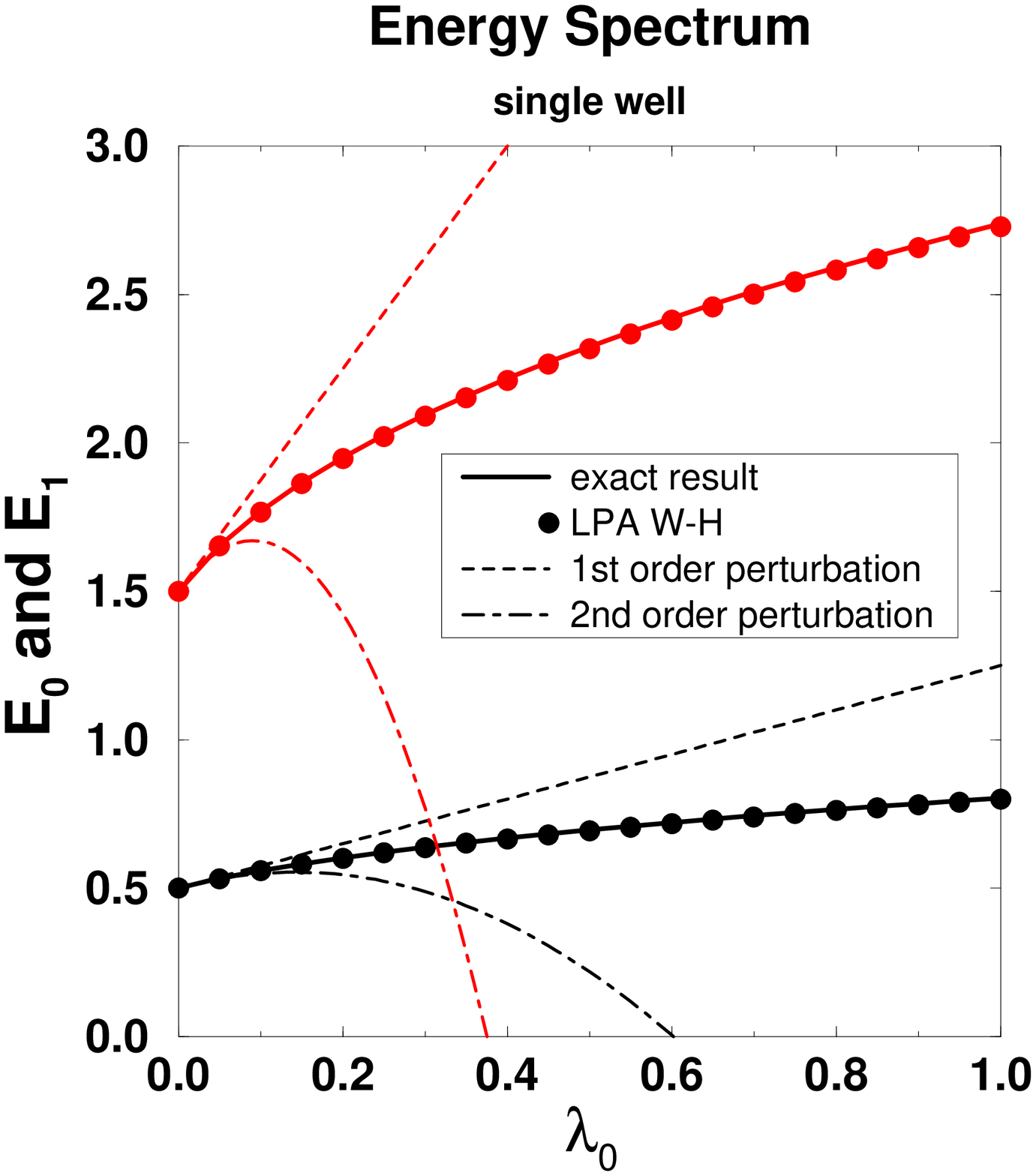}
\vspace{-3mm}
 \caption{Energy spectrum.}
 \label{fig:sspe}    
 }
\end{figure}

Now we proceed to analyze quantum mechanical anharmonic
oscillators and double well systems.
First, we consider a symmetric single-well potential, 
\begin{eqnarray}
V_0(x) =~~\lambda_0 x^4+\frac{1}{2} x^2.\label{(30)}
\end{eqnarray}
Our interest is to compare our
NPRG results with the perturbative
series. 
First, the LPA W-H equation (\ref{(16)}) is solved numerically,
and we thereby obtain an  
effective potential $V_{\rm eff}$.
The flow of $V_{\Lambda}$ is shown in Figure.\ref{fig:spote}.
Quantum corrections raise the potential and
make its slope steeper.
In Figure.\ref{fig:sspe}, we display the energy spectrum calculated 
with the relations (\ref{(20)}) and (\ref{(23)}).
We refer to the results obtained by 
a numerical analysis of the Schr\"odinger equation as
the ``exact results.''
\par
\begin{figure}[htb]
\hspace{0mm}
 \parbox{65mm}{
 \epsfxsize=65mm 
 \epsfysize=65mm
  \leavevmode
\epsfbox{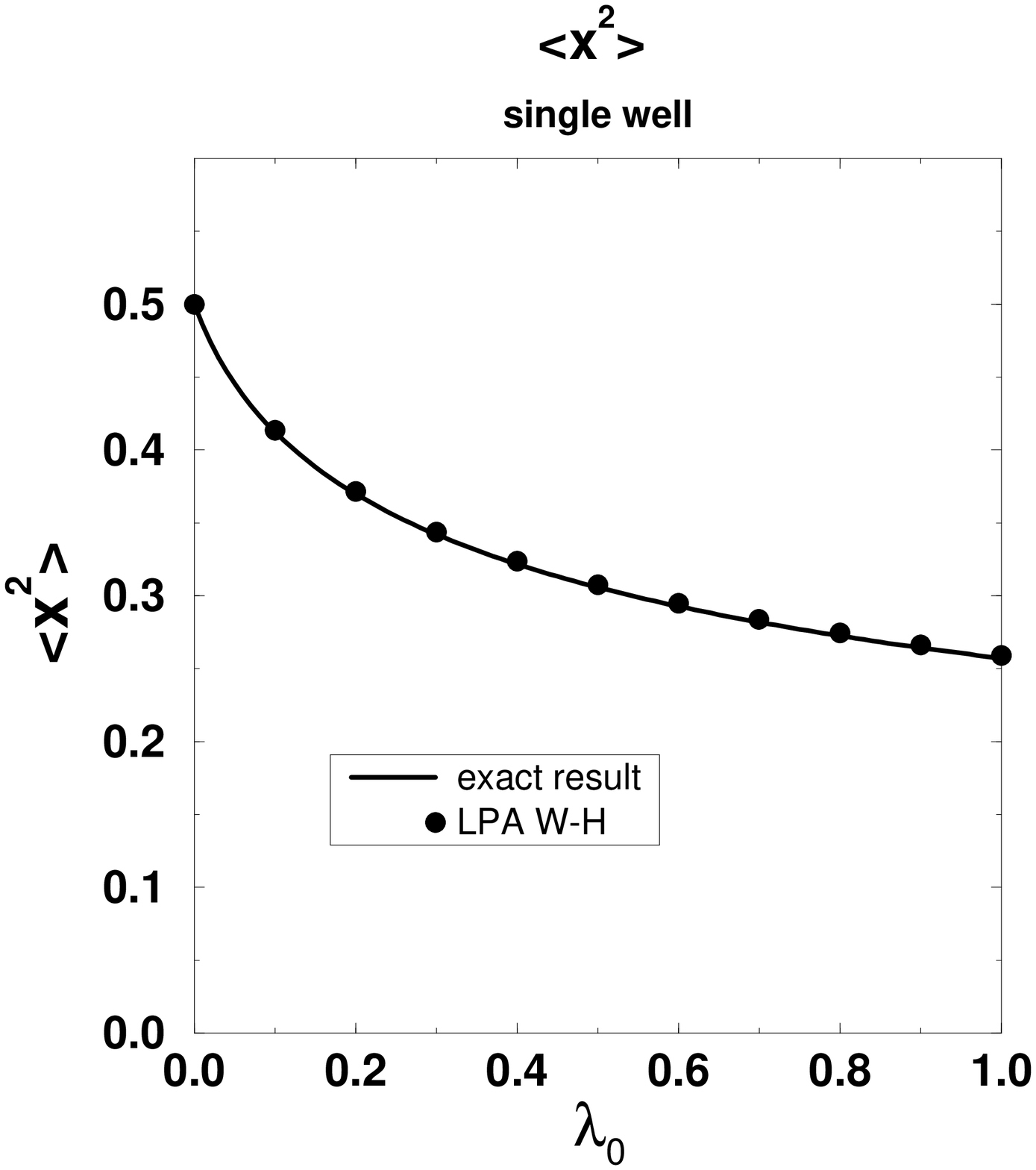}
\vspace{-3mm}
 \caption{$M_2 =
\langle \Omega |~\hat{x}^2| \Omega \rangle$.}
 \label{fig:sx2}
 }
\hspace{0mm} 
\parbox{65mm}{
 \epsfxsize=65mm 
 \epsfysize=65mm
 \leavevmode
\epsfbox{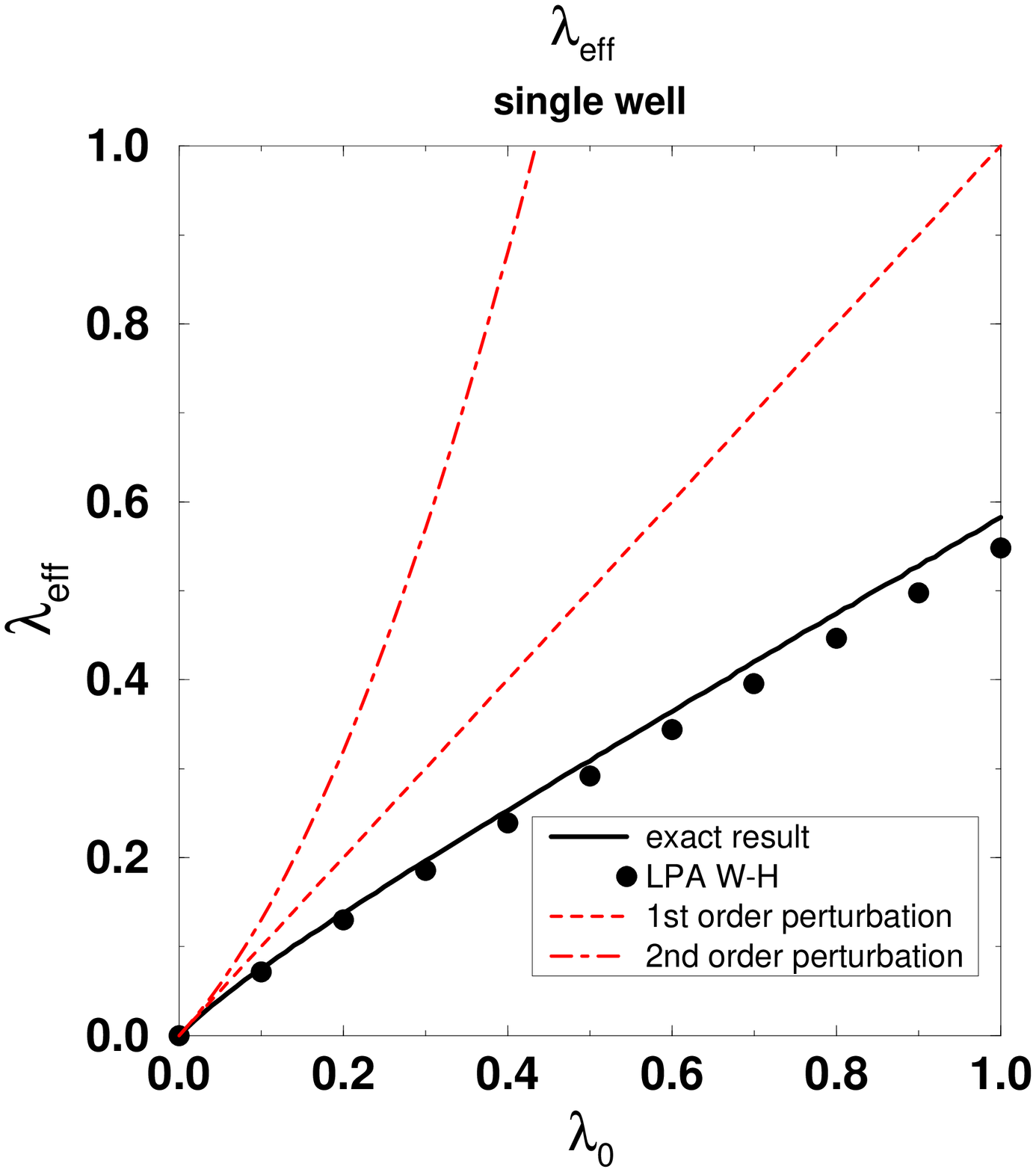}
\vspace{-3mm}
 \caption{Effective coupling  $\lambda _{\rm eff}$.}
 \label{fig:seffLa}    
 }
\end{figure}
The perturbative series of $E_n$ is the asymptotic series 
\begin{eqnarray}
E_n\!=\![n+\frac{1}{2}]+\frac{3}{4}\lambda_0 [2n^2+2n+1]
-\frac{1}{8}\lambda_0 ^2 [34n^3+51n^2+59n+21]+\cdots .~~~~ \label{(31)}
\end{eqnarray}
It diverges even in the weak coupling region.
Note that the Borel resummation of the perturbative series works well 
in this case, and gives quantitatively good values. 
However, even in the lowest order approximation (LPA), 
the W-H equation can evaluate the energy spectrum almost perfectly. 
Therefore, we conclude that the NPRG does sum up all orders of 
the perturbative series in the correct manner. 
\par
We also display the two-point function 
$M_2 =\langle \Omega |~\hat{x}^2| \Omega \rangle$ 
in Figure.\ref{fig:sx2} 
and the effective four-point coupling constant $\lambda_{\rm eff} $
in Figure.\ref{fig:seffLa}.
In both of these cases as well, the NPRG results give an 
almost perfect fit.
\subsection{Symmetric double-well potential}
\begin{figure}[htb]       % 
\hspace{0mm}
 \parbox{65mm}{
 \epsfxsize=65mm      % 	
 \epsfysize=65mm
  \leavevmode
\epsfbox{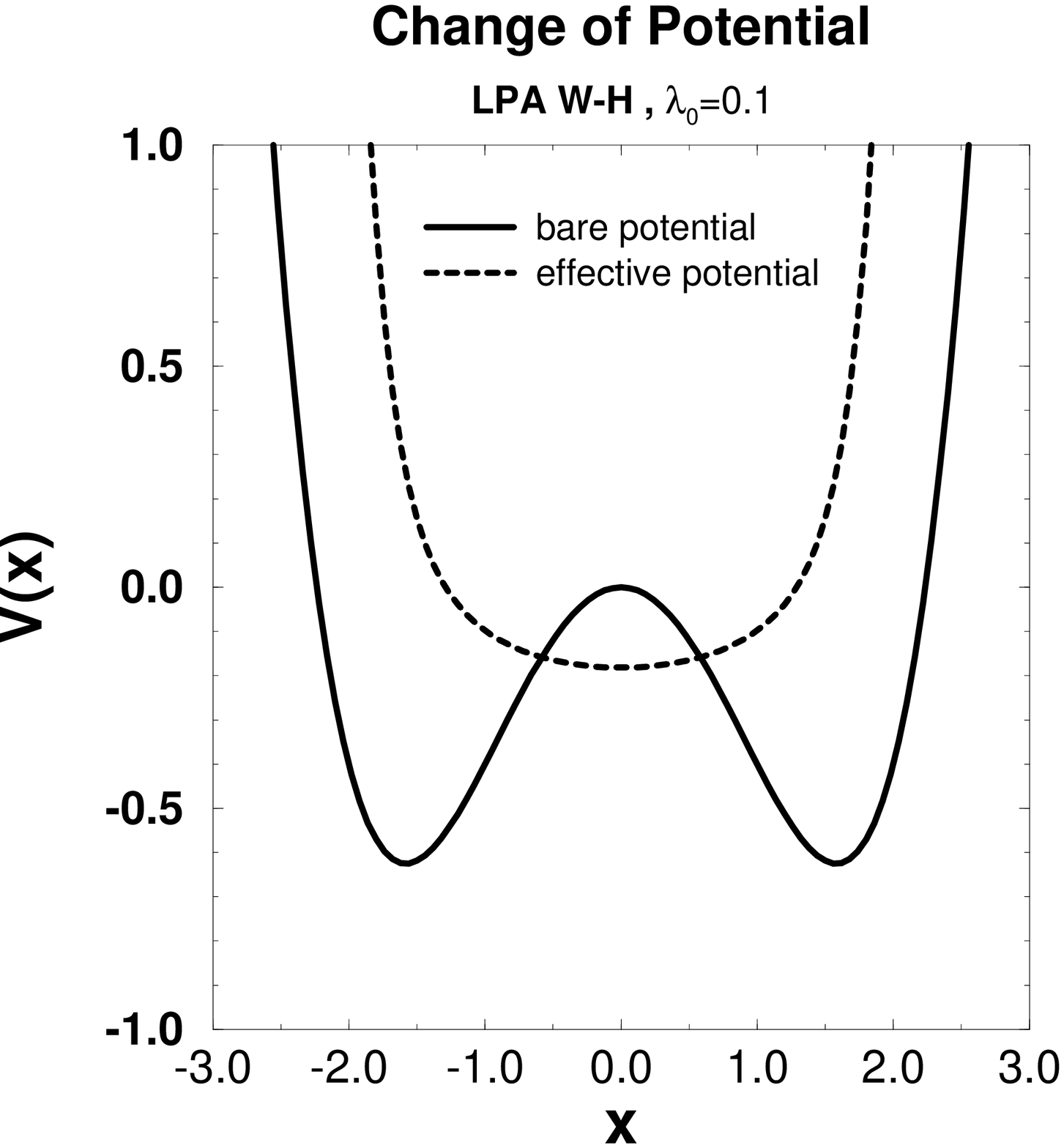}
\vspace{-3mm}
 \caption{Flow of the potential.}
 \label{fig:spote3}        %
 }
\hspace{0mm} 
\parbox{65mm}{
 \epsfxsize=65mm      % 
 \epsfysize=65mm
 \leavevmode
\epsfbox{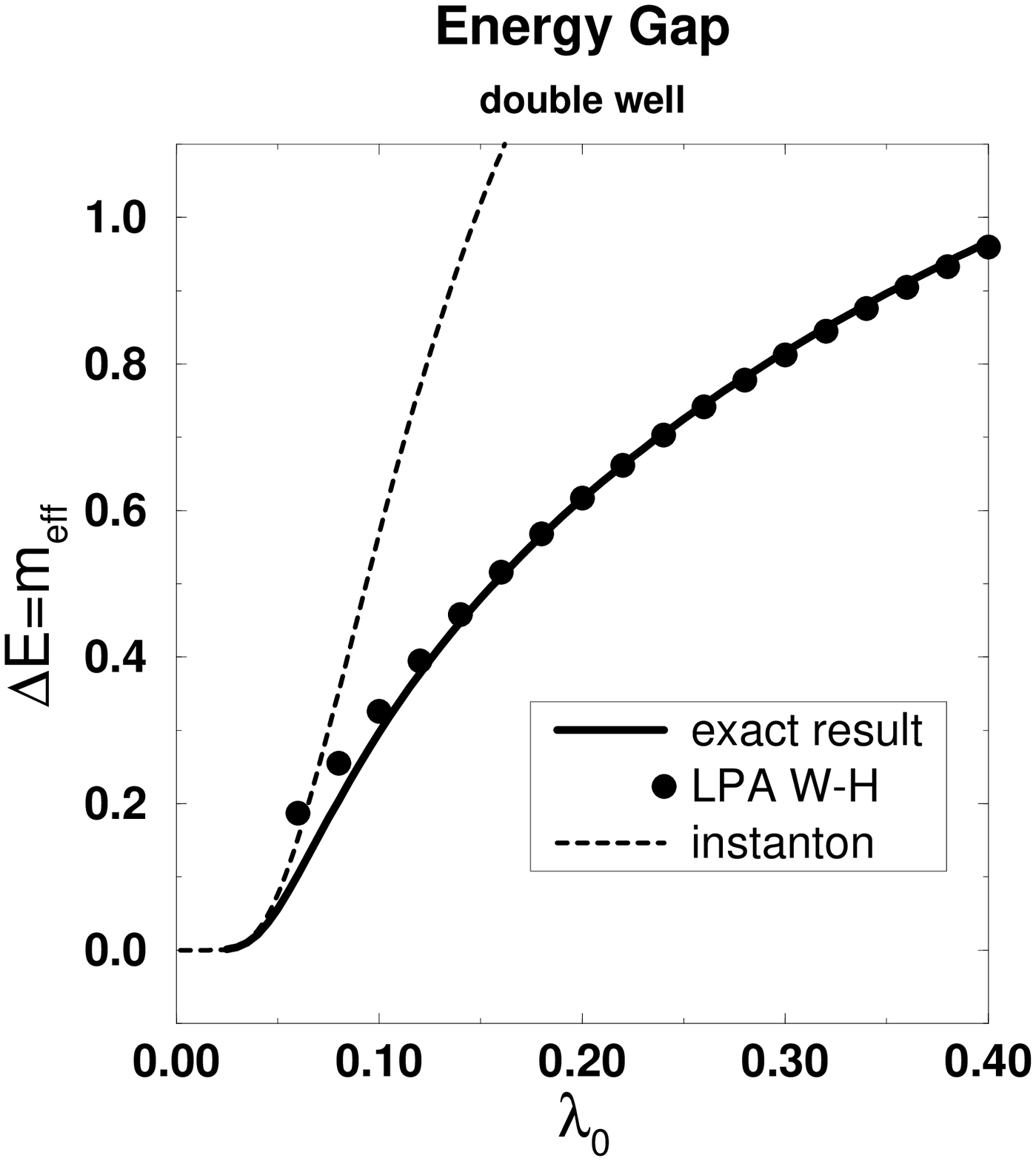}
\vspace{-3mm}
 \caption{Energy gap estimates.}        %
 \label{fig:smass}        %γκ
 }
\end{figure}

Next, we consider the $Z_2$-symmetric $x\leftrightarrow -x$ 
double-well potential,
\begin{eqnarray}
V_0(x) =\lambda_0 x^4-\frac{1}{2}x^2. \label{(32)}
\end{eqnarray}
In quantum mechanical systems, this $Z_2$ symmetry never 
breaks spontaneously,
because the $x$ mode tunnels 
through the potential barrier, and 
the ground state is uniquely realized. 
In fact, in the NPRG evolution of the effective potential, 
the initial double-well potential finally becomes 
a single well, and an energy gap (effective mass)
arises (Figure.\ref{fig:spote3}).
\par
In this system, there is no well-defined perturbation theory. 
A standard technique to obtain the energy gap $\Delta E$ is 
the dilute gas instanton calculation. 
This is a semi-classical method based on 
the one-instanton solutions
\begin{eqnarray}
x_{\rm cl}(\tau)=\pm \frac{1}{2\sqrt{\lambda_0}} 
\tanh \frac{(\tau-\tau_0)}{\sqrt{2}}. \label{(33)}
\end{eqnarray}
The one-instanton contribution to the partition function $Z$ is
\begin{eqnarray}
Z\simeq Te^{-S\bka{x_{\rm cl}}}
    \sqrt{\frac{S\bka{x_{\rm cl}}}{2\pi}}
    \mka{ \frac{
     \det ^{\prime}\ska{\left.
              \frac{\delta ^2 S}{\delta x^2}
              \right|_{x=x_{\rm cl}} }}{     
     \det\ska{\left.
              \frac{\delta ^2 S}{\delta x^2}
              \right|_{x=\pm \frac{1}{2\sqrt{\lambda_0}}} }
         }
         }^{-1/2} \equiv T\frac{\Delta _0}{2}, \label{(34)}
\end{eqnarray} 
where $T$ is an imaginary time volume.
Assuming that instantons do not interact with each other,
we can evaluate the multi-instanton contribution to $Z$ 
(the dilute gas instanton approximation), and we obtain 
the energy gap 
\begin{eqnarray}
\Delta E=\Delta _0=2\sqrt{\frac{2\sqrt{2}}{\pi \lambda_0}}
e^{-\frac{1}{3\sqrt{2}\lambda_0}}, \label{(35)}
\end{eqnarray}
which has the structure of an essential singularity originating 
from the one-instanton action.
The singularity coefficient obtained from the instanton method
is known to be exact in the vanishing $\lambda_0$ limit.\cite{simon}
\par
\vspace{3mm}
\par
\begin{figure}[htb]
\hspace{0mm}
 \parbox{65mm}{
 \epsfxsize=65mm 
 \epsfysize=65mm
  \leavevmode
\epsfbox{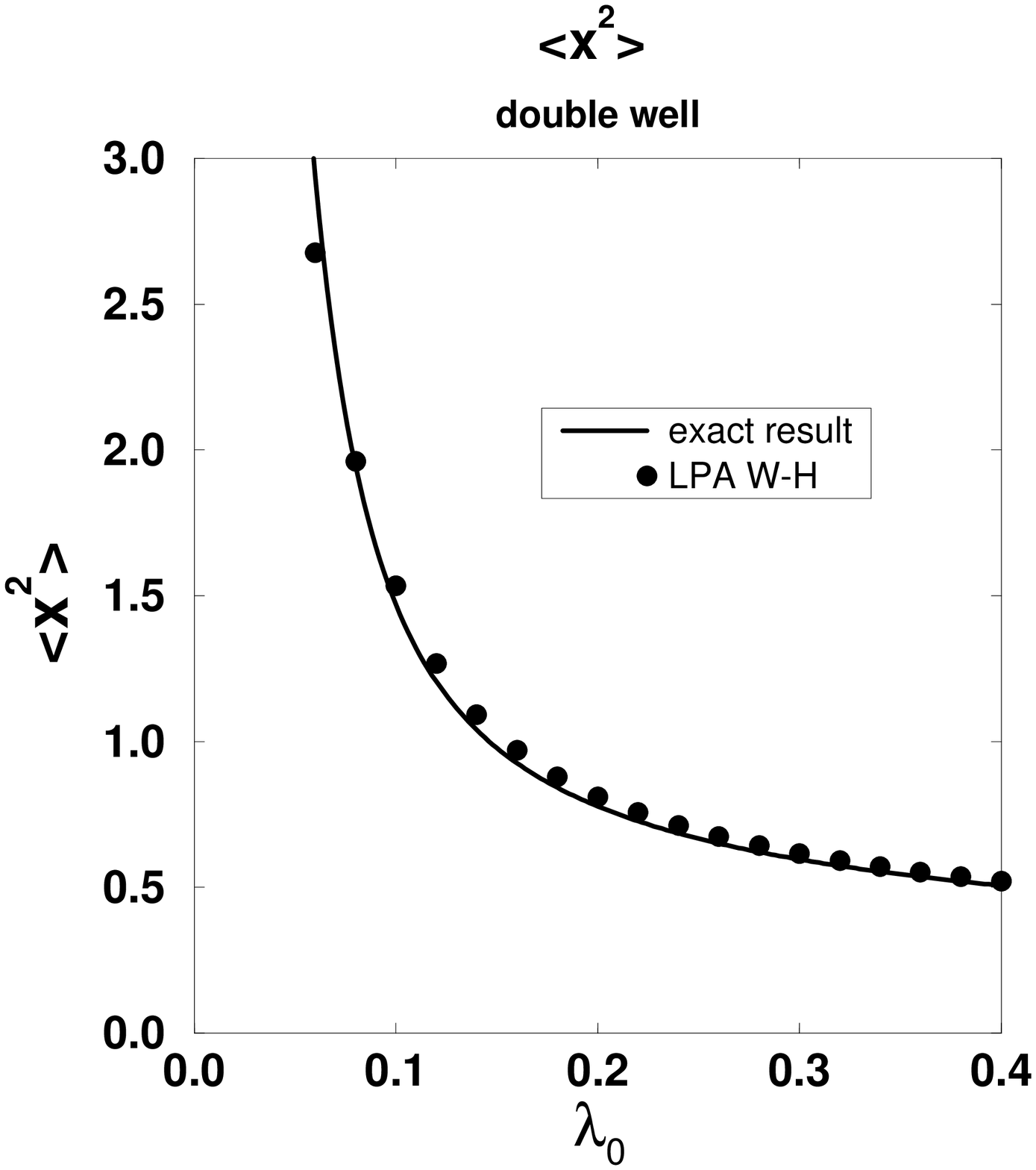}
\vspace{-3mm}
 \caption{$M_2 =
\langle \Omega |~\hat{x}^2| \Omega \rangle$.}
 \label{fig:dx2}
 }
\hspace{0mm} 
\parbox{65mm}{
 \epsfxsize=65mm 
 \epsfysize=65mm
 \leavevmode
\epsfbox{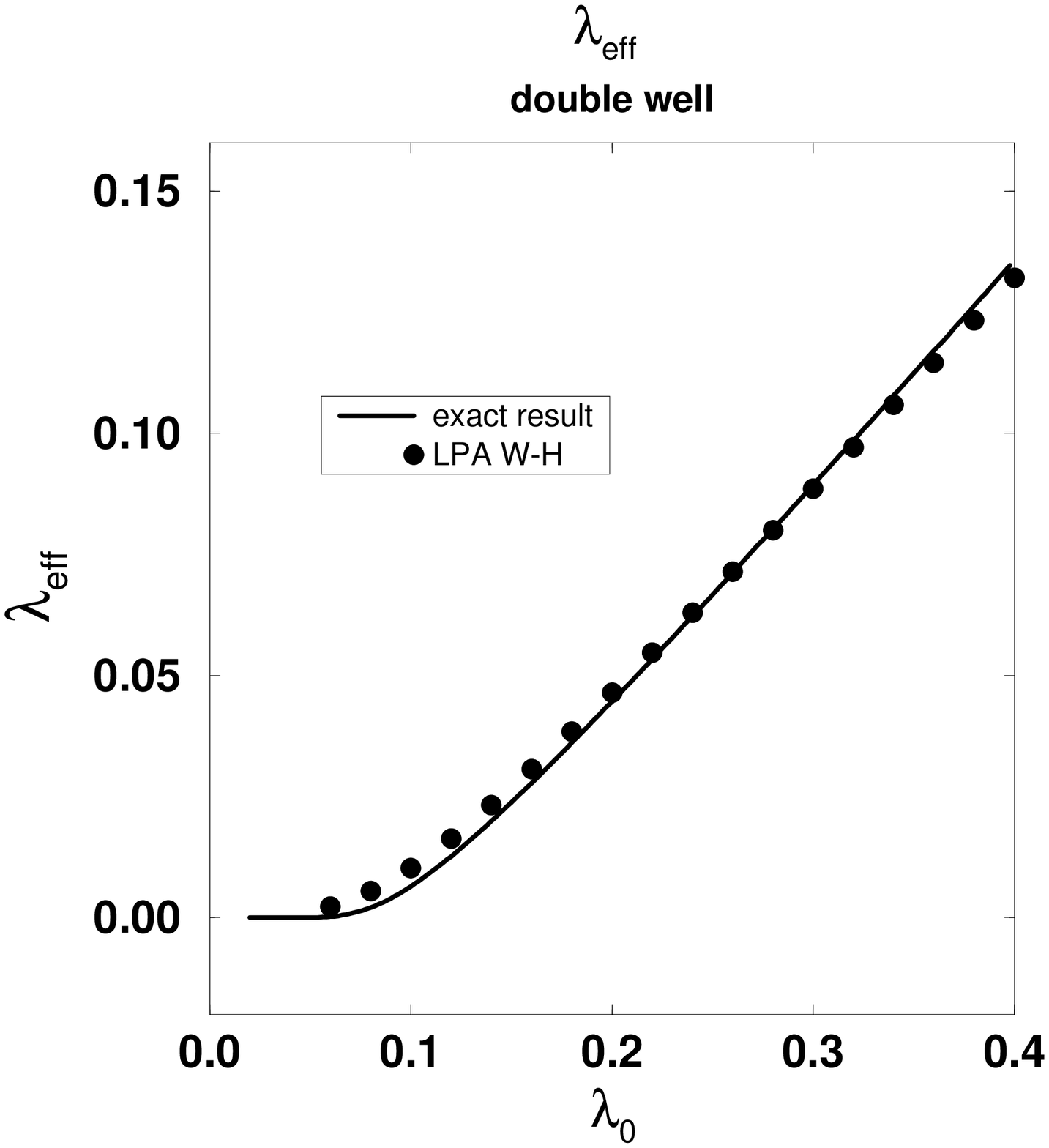}
\vspace{-3mm}
 \caption{Effective coupling $\lambda _{\rm eff}$.}
 \label{fig:deffLa}    
 }
\end{figure}
\par
In Figure.\ref{fig:smass} we display the energy gap 
evaluated using various methods.
The NPRG results are very good in the strong coupling region, 
while the perturbation cannot be applied in this double-well system, 
and the dilute gas instanton method is not at all effective, 
because it is valid only in the very weak coupling region.\footnote{
It has long been known that the strong coupling expansion has 
a finite radius of convergence. 
Recently, variational perturbation theory has become highly developed, 
and very accurate results have been obtained.\cite{kl,gks}
The region of coupling constant values in which these approaches are good
is estimated as  $\lambda_0 \gsim 0.08$, which is 
almost coincident with the reliable region for our method.
To elucidate the correspondence between the NPRG method and 
this improved perturbation theory is interesting.
}
Therefore, the NPRG method should provide a
powerful tool for the analysis of tunneling, at least in such
regions.  
However, our NPRG results deviate from the exact values as $\lambda _0 \to
0$, which corresponds to a very deep well. Because the $\beta$-function
becomes singular in this region, the NPRG results become unreliable. We
believe that the cause of the difficulty is the
LPA approximation scheme that we adopt. 
It is important to note that the respective coupling regions 
in which the LPA W-H equation and the dilute gas instanton work well 
are separated, 
and therefore these two methods should be regarded 
as complementary.\cite{ahtt,ggm}
\par
We display the two-point function 
$M_2 =\langle \Omega |~\hat{x}^2| \Omega \rangle$ 
in Figure.\ref{fig:dx2} 
and the effective four-point coupling $\lambda_{\rm eff} $
in Figure.\ref{fig:deffLa}.
As in the case of the energy gap, 
the NPRG results are excellent here, except in  
the extremely weak coupling region. 
\subsection{Flow diagrams}
\begin{figure}[htb]       % 
\hspace{0mm}
 \parbox{65mm}{
 \epsfxsize=65mm      % 	
 \epsfysize=65mm
  \leavevmode
\epsfbox{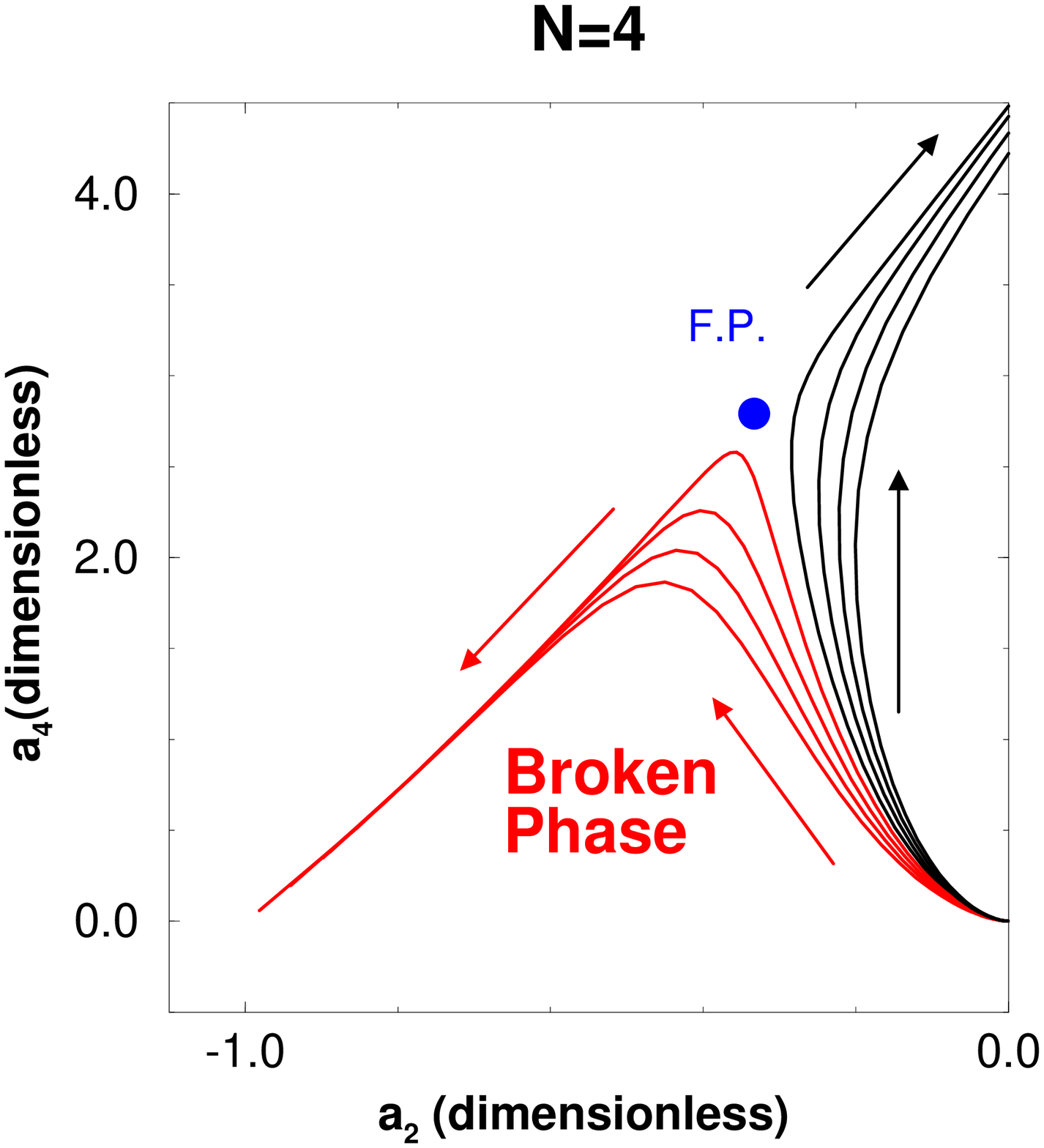}
\vspace{-3mm}
\caption{$N=4$ truncation.}
\label{fig:sWHflowN4}        %
 }
\hspace{0mm} 
\parbox{65mm}{
 \epsfxsize=65mm      % 
 \epsfysize=65mm
 \leavevmode
\epsfbox{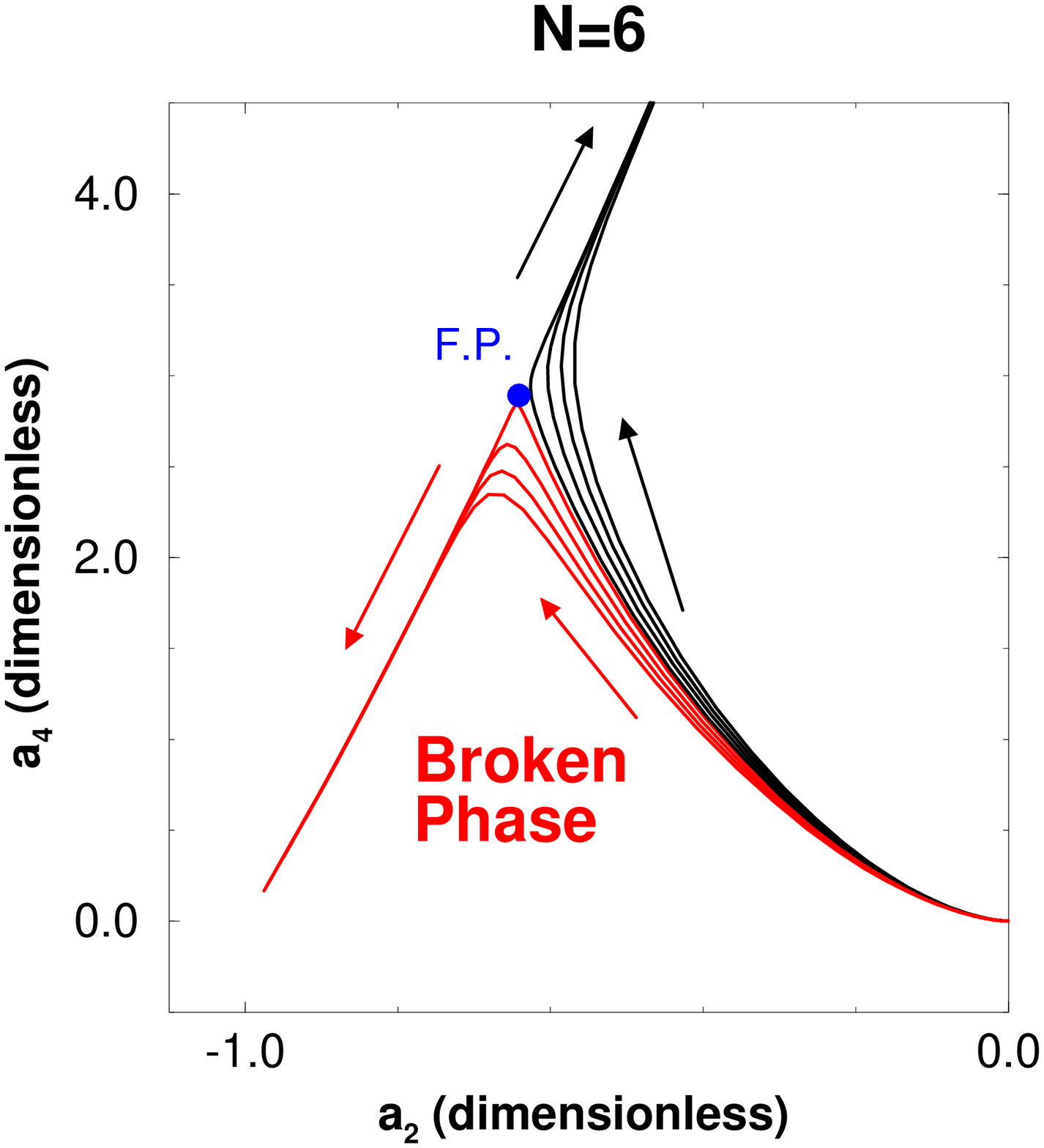}
\vspace{-3mm}
\caption{$N=6$ truncation.}        %
\label{fig:sWHflowN6}        %
 }
\end{figure}

\begin{figure}[htb]       % 
\hspace{0mm}
 \parbox{65mm}{
 \epsfxsize=65mm      % 	
 \epsfysize=65mm
  \leavevmode
\epsfbox{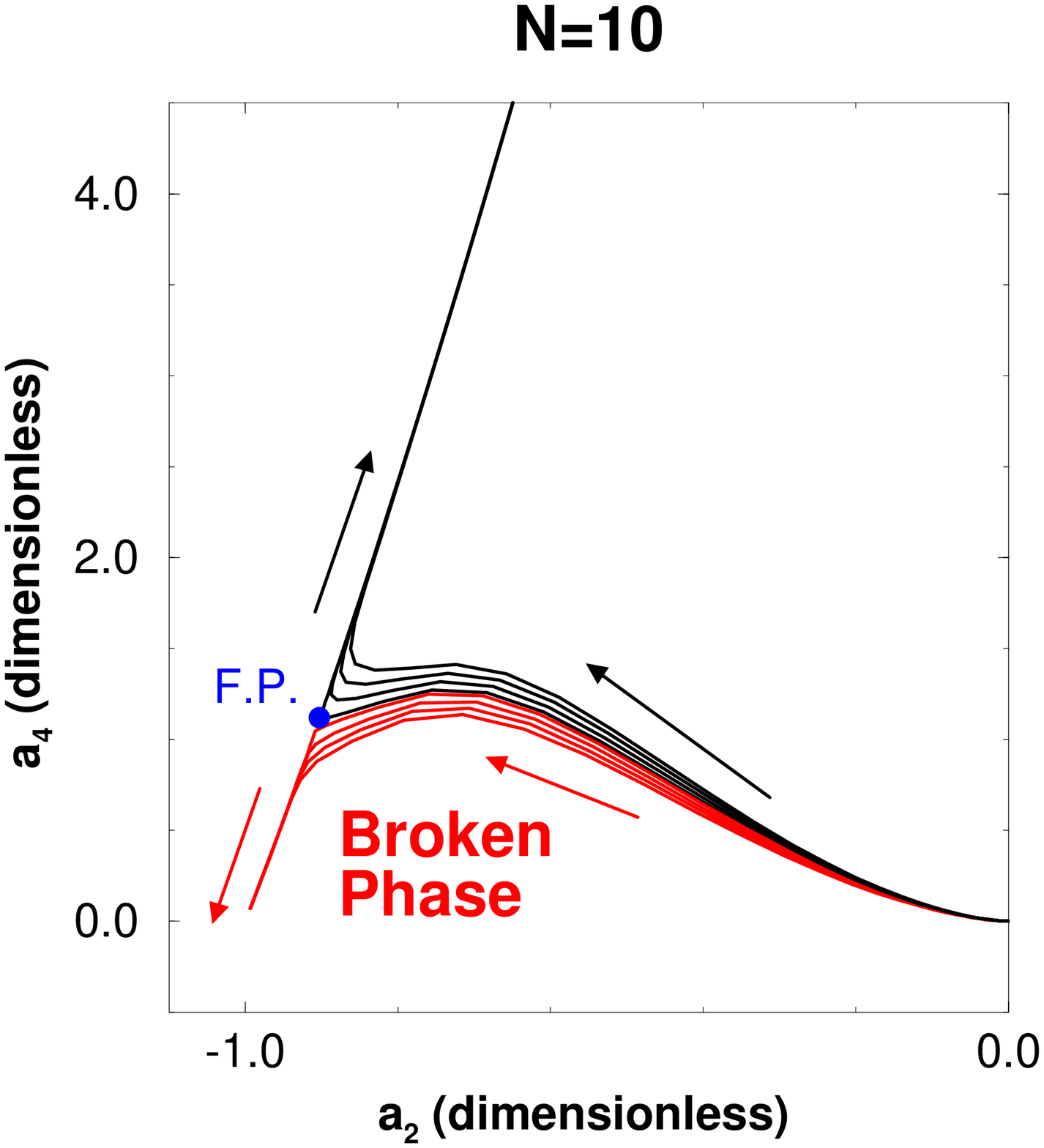}
\vspace{-3mm}
\caption{$N=10$ truncation.}
\label{fig:sWHflowN10}        %
 }
\hspace{0mm} 
\parbox{65mm}{
 \epsfxsize=65mm      % 
 \epsfysize=65mm
 \leavevmode
\epsfbox{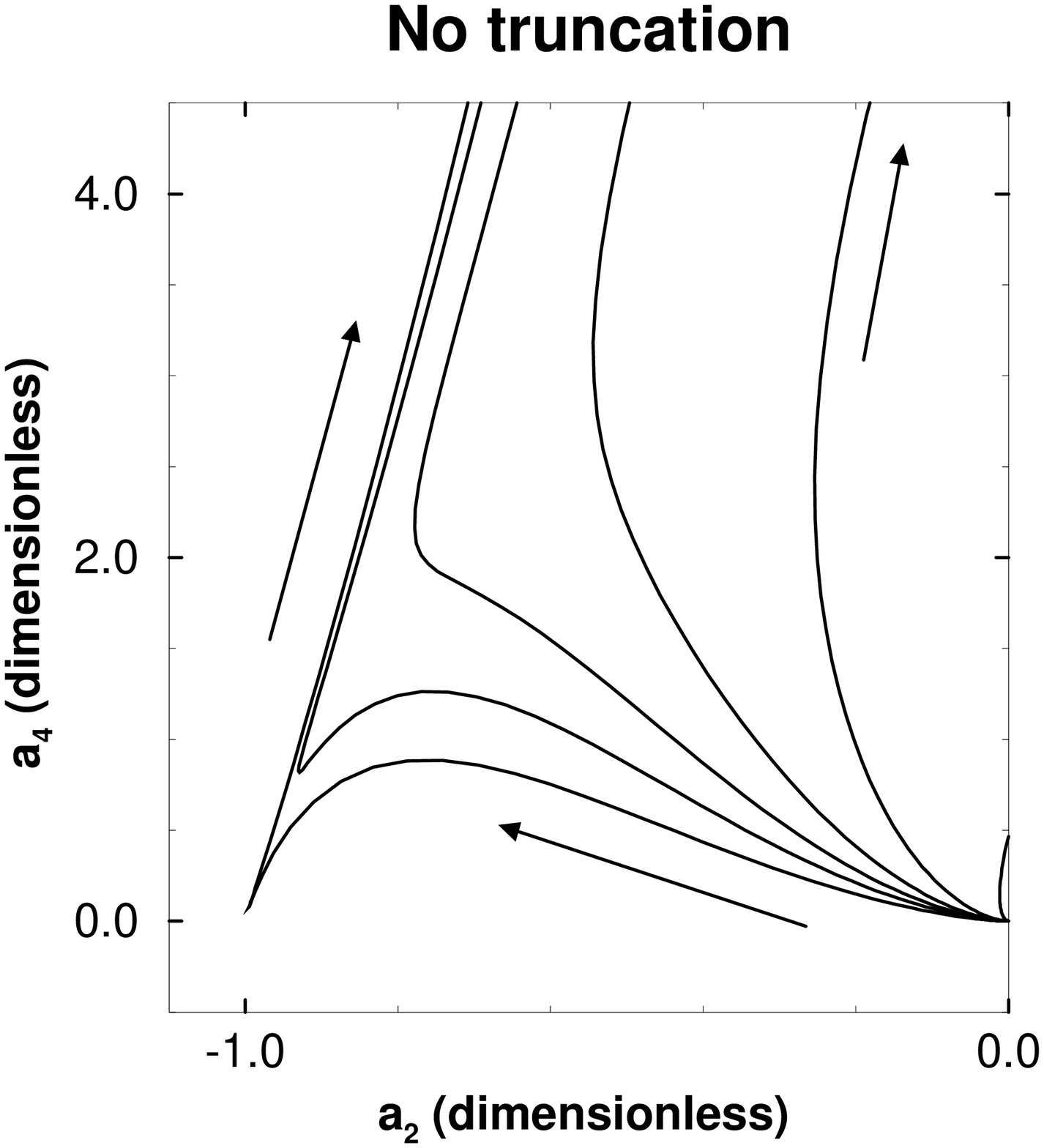}
\vspace{-3mm}
\caption{No truncation.}        %
\label{fig:sflow24}        %
 }
\end{figure}

We now more carefully consider the difficulty  
arising in the weak coupling region for the double-well potential.
We employ the operator expansion (\ref{(10)}) and
investigate the flows of the dimensionless coupling constants 
$\hat{a}_n\equiv a_n \Lambda^{-\frac{n+2}{2}}$. 
The flow diagrams elucidate the phase structure of the system.
We display the flow diagrams for the $N=4,6,10$ truncated potentials 
(Figure.\ref{fig:sWHflowN4},  Figure.\ref{fig:sWHflowN6},
Figure.\ref{fig:sWHflowN10})
and for the potential without an 
operator expansion (Figure.\ref{fig:sflow24}). 
\par
These flow diagrams reveal that the phase structure of the theory
with a truncated potential ($N=4,6,10$) is  
somewhat strange.
As mentioned above, there is no spontaneous symmetry breaking 
in these quantum mechanical systems.
When we truncate the potential as in (\ref{(10)}), 
there appears a non-actual fixed point and false broken phases
(Figures \ref{fig:sWHflowN4}, \ref{fig:sWHflowN6} 
and \ref{fig:sWHflowN10}).   
The flow starting from the weak coupling region 
($\lambda_{0} \to 0$ i.e. $\hat{a_4}\to 0$) 
tends to be captured by the false broken phase, and 
we cannot obtain the correct result $m^2_{\rm eff}>0$.
The region of the false broken phase 
becomes smaller as $N$ becomes larger, and then
for the LPA exact (no truncation) calculation, 
the false broken phase disappears (Figure.\ref{fig:sflow24}). 
However, even in the no truncation case,
we cannot obtain reliable results for 
the flows that start from the weak coupling region, 
because singular behavior of the flow 
in the region near $\hat{a}_2=-1$ 
leads to large numerical errors.  
The results for the LPA W-H equation in Figure.\ref{fig:smass}
were obtained from numerical integration of the partial 
differential equation
without any truncation.\cite{kt}
\begin{wrapfigure}{r}{6.6cm}
\epsfysize=69mm
\epsfxsize=69mm
\leavevmode
\epsfbox{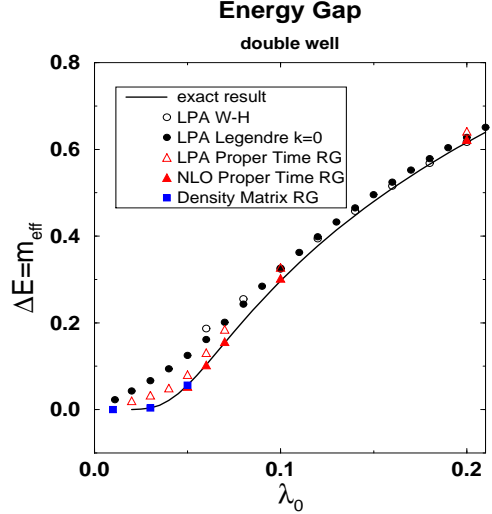}
\vspace{-4mm}
\caption{Various RG results.}
 \label{fig:various}
\end{wrapfigure}
\subsection{Other methods}
The NPRG equation we employ here is that with the local potential
approximation. The results in the weak coupling region can be
improved by upgrading the approximation.
The LPA is the lowest order of the derivative expansion, 
and a higher-order 
calculation can be carried out.\cite{morris}   
In this quantum mechanical system, the second-order calculation 
of the Legendre flow equation does not improve 
the weak coupling results.\cite{ah}
However, an analysis using the proper time renormalization group
improves the LPA results considerably.\cite{za}
Also, although it differs from the NPRG methods in its formulation,
the density matrix renormalization group is useful 
for this system.\cite{mdsn}
We exhibit in Figure.\ref{fig:various} the results 
for the energy gap obtained 
with various renormalization group approaches.\footnote{
As for general non-perturbative methods, 
the auxiliary field method works very well
both in the weak coupling and strong coupling regions.\cite{taro}
Also, various improved perturbation theories have been applied to the 
anharmonic oscillator and the double well system, giving similar
results.\cite{kl,gks,kuni1,kuni2}
}
\subsection{Asymmetric double-well potential}
We proceed to consider the $Z_2$-asymmetric double-well potential
\begin{eqnarray}
V_0(x) =~~\lambda_0 x^4-\frac{1}{2} x^2+h_0 x, \label{(36)}
\end{eqnarray}
where the linear term $h_0x$ breaks the $Z_2$ symmetry explicitly. 
In this system there are a stable minimum and an unstable minimum.
\par
How do we deal with the effect of such an asymmetric term?
The NPRG method can treat this system in a manner that is quite similar 
to that for the symmetric system;
it just changes the initial potential,
while the LPA W-H equation does not change.
Furthermore, when we apply the operator expansion (\ref{(10)}),
the situation becomes even simpler. 
The additional $h_0x$ term does not affect the running of other 
coupling constants, because the term 
\begin{equation}
\int d\tau \!~
h_0x(\tau)=h_0 x(E=0), \label{(37)}
\end{equation}
consists entirely of the zero energy mode, 
and generates no quantum corrections. 
Therefore, the NPRG equations for the coupling constants are 
the same as those in the symmetric case.
\par
By contrast, the standard instanton method 
cannot be applied to such an asymmetric system, because 
the term 
$\left.\frac{\delta^2 S}{\delta x\delta x}\right|_{x=x_{\rm cl}}$
in (\ref{(34)}) has a negative eigenvalue in this case.
For actually unstable systems, this negative eigenvalue
is converted to a decay rate for the system.
This is a typical prescription for the `bounce solution' calculation.
However, in the case of the bare potential (\ref{(32)}),
the true vacuum of the system is stable.
The existence of a negative eigenvalue in actually stable systems
is known as the problem of a fake instability.
To overcome this problem, the valley method 
has been developed recently.\cite{ao1,ao2}
It is a generalization of the instanton method 
that is based on the
valley structure in the configuration space.
\par
\begin{figure}[htb]  
\hspace{0mm}
 \parbox{65mm}{
 \epsfxsize=65mm     
 \epsfysize=65mm
  \leavevmode
\epsfbox{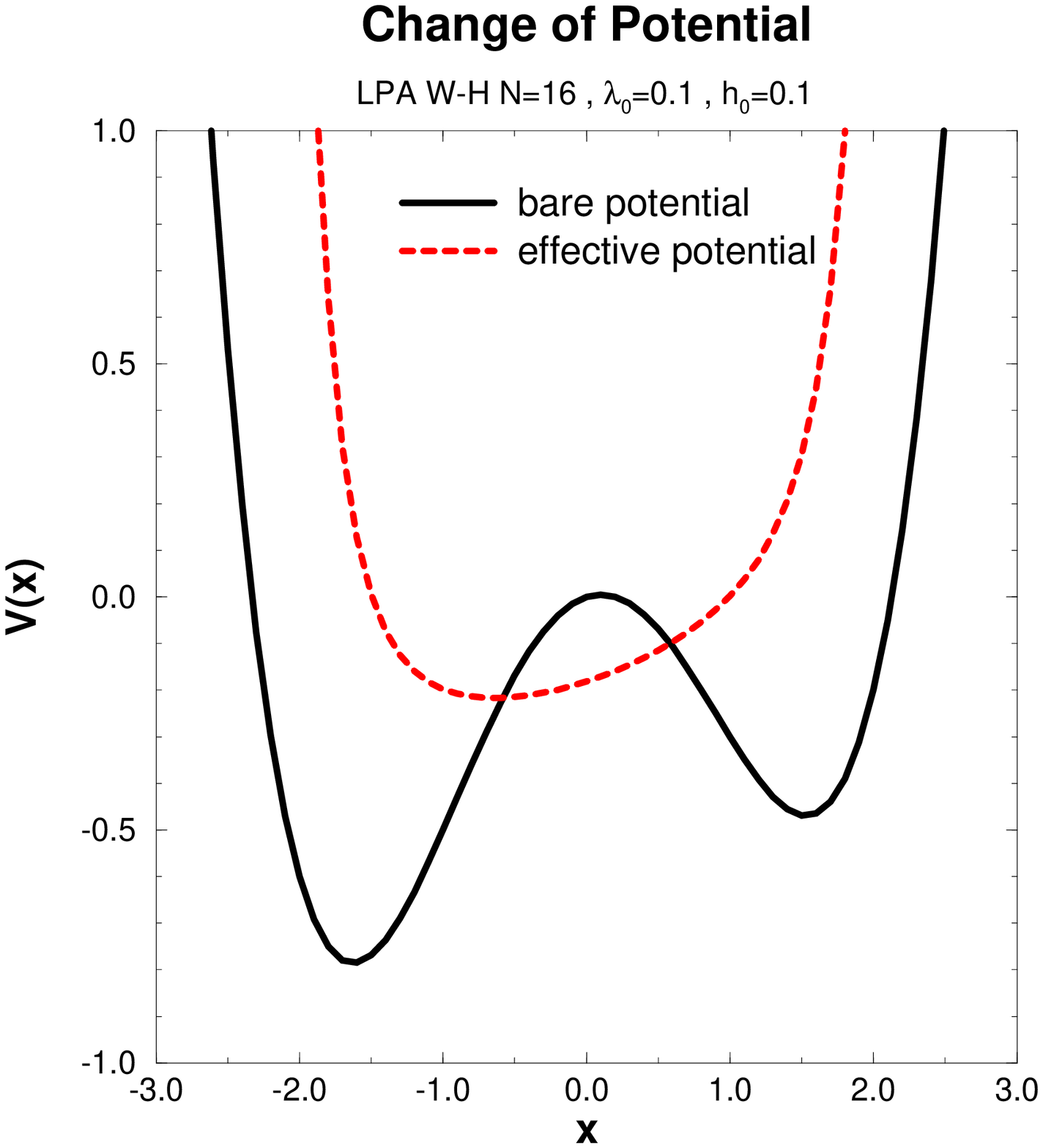}
\vspace{-3mm}
 \caption{Flow of the potential.}
 \label{fig:adpote}   
 }
\hspace{0mm} 
\parbox{65mm}{
 \epsfxsize=65mm     
 \epsfysize=65mm
 \leavevmode
\epsfbox{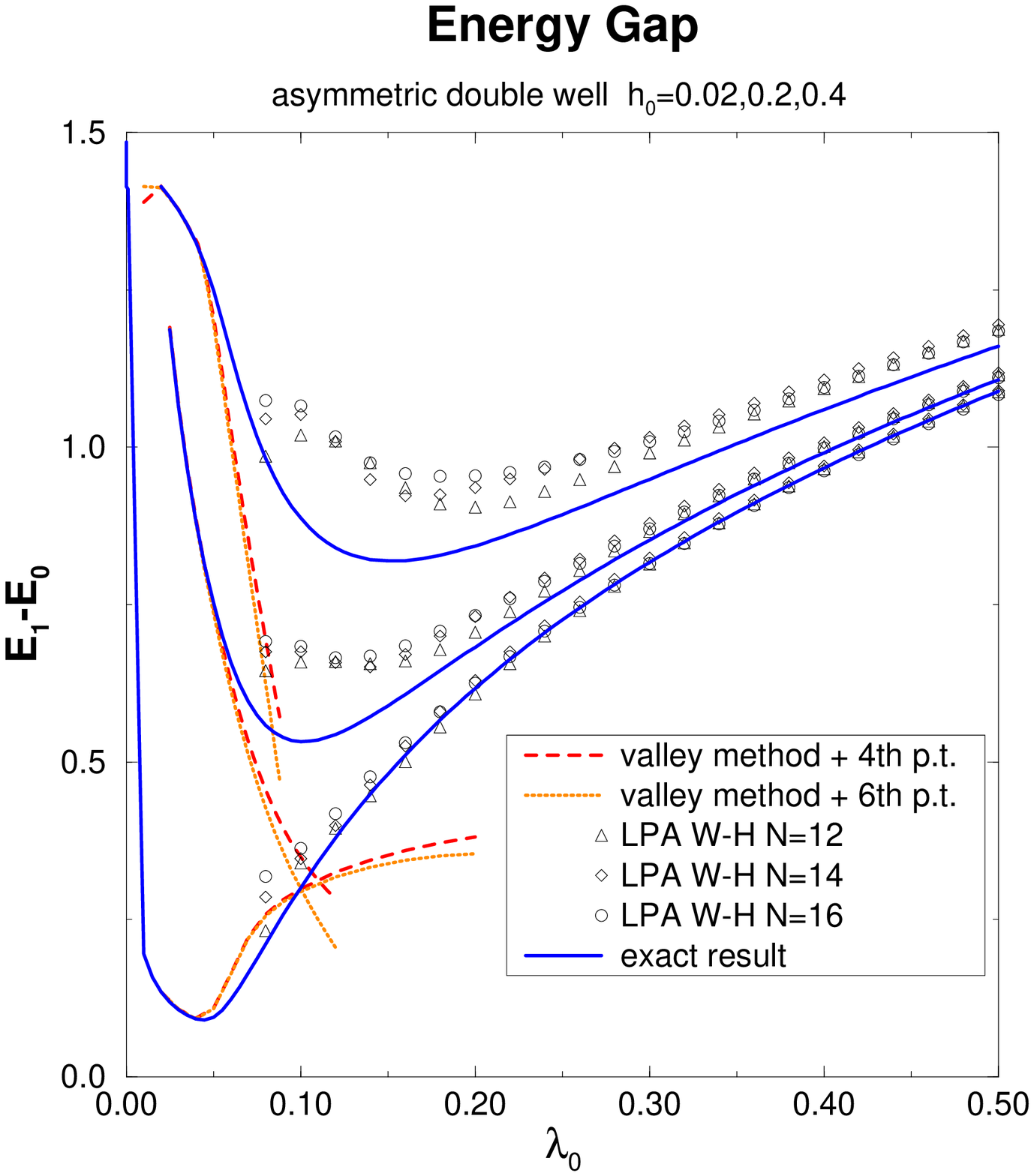}
\vspace{-3mm}
 \caption{Energy gap estimates.}
 \label{fig:adspe}   
 }
\end{figure}

As shown in Figure.\ref{fig:adpote},
an asymmetric bare potential leads to an asymmetric effective potential. 
We show in Figure.\ref{fig:adspe} results for the energy gap in the
cases of  
three values of $h_0$, from bottom to top, $h_0=0.02,\!~0.2,\!~0.4$.
For any value of $h_0$, in the $\lambda_0\to 0$ limit,  
$\Delta E$ approaches $\sqrt{2}$.
This is because in this limit the asymmetric double well approaches 
a single well.
We employ the operator expansion and give the truncation 
$N=12,14,16$ results.
We also plot the results obtained from the valley method 
with fourth and sixth order perturbations.\cite{ao4}
A complementary relation between the NPRG and the valley method
is observed, just as in the case of the symmetric potential.
\par
As mentioned above, since in the $\lambda_0\to 0$ limit 
the potential approaches a single well, 
if we carry out the operator expansion 
at the potential minimum $x=x_{\rm min}$,
the NPRG equations never become singular even in the 
$\lambda_0\to 0$ region,
and we obtain $m_{\rm eff}\simeq\sqrt{2}$.
We use this technique for analysis of SUSY QM in the next section.
\par
\begin{figure}[htb]
\hspace{0mm}
 \parbox{65mm}{
 \epsfxsize=65mm 
 \epsfysize=65mm
  \leavevmode
\epsfbox{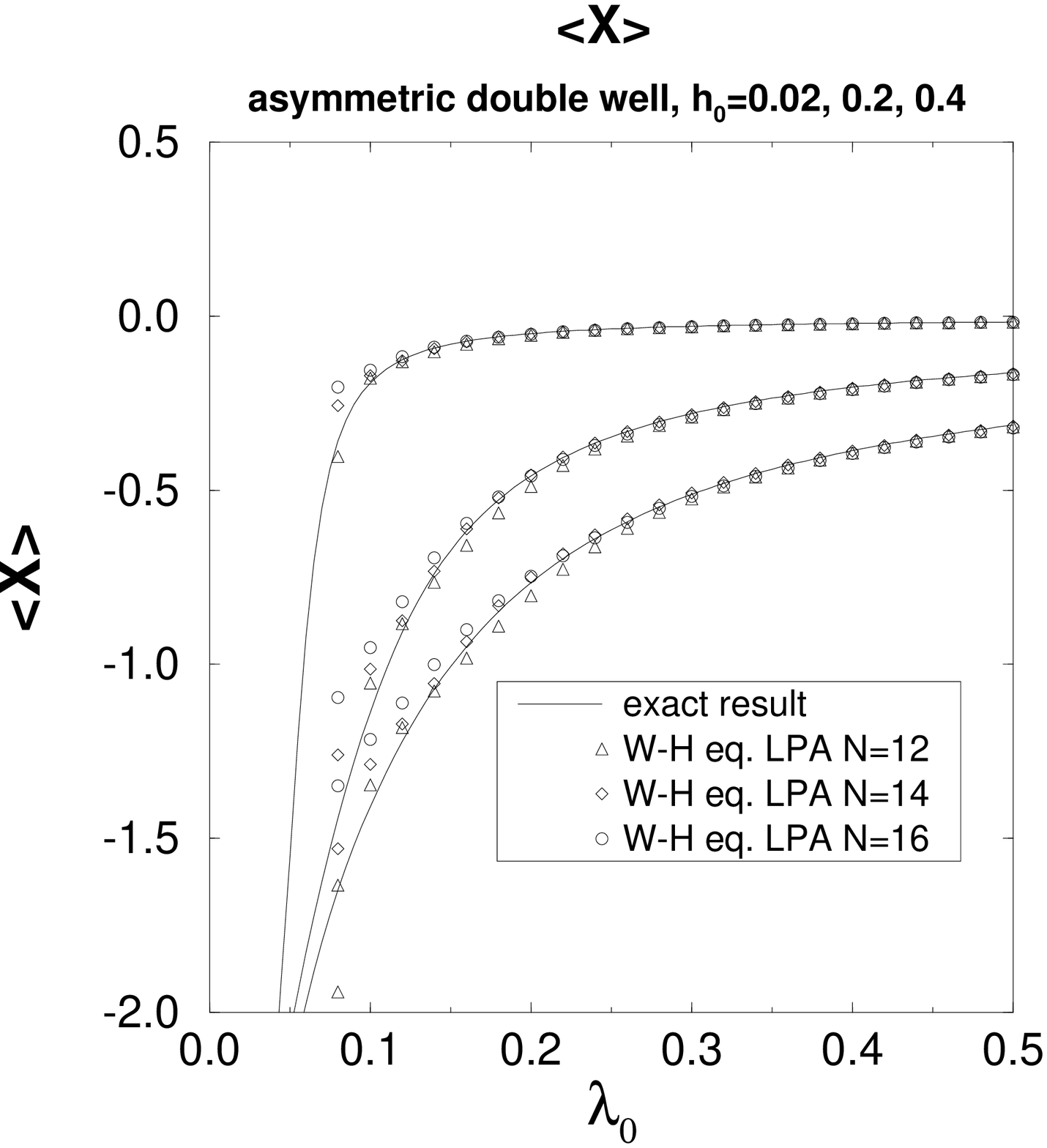}
\vspace{-3mm}
 \caption{$M_1 =
\langle \Omega |\hat{x}| \Omega \rangle$
}
 \label{fig:asymXN}
 }
\hspace{0mm} 
\parbox{65mm}{
 \epsfxsize=65mm 
 \epsfysize=65mm
 \leavevmode
\epsfbox{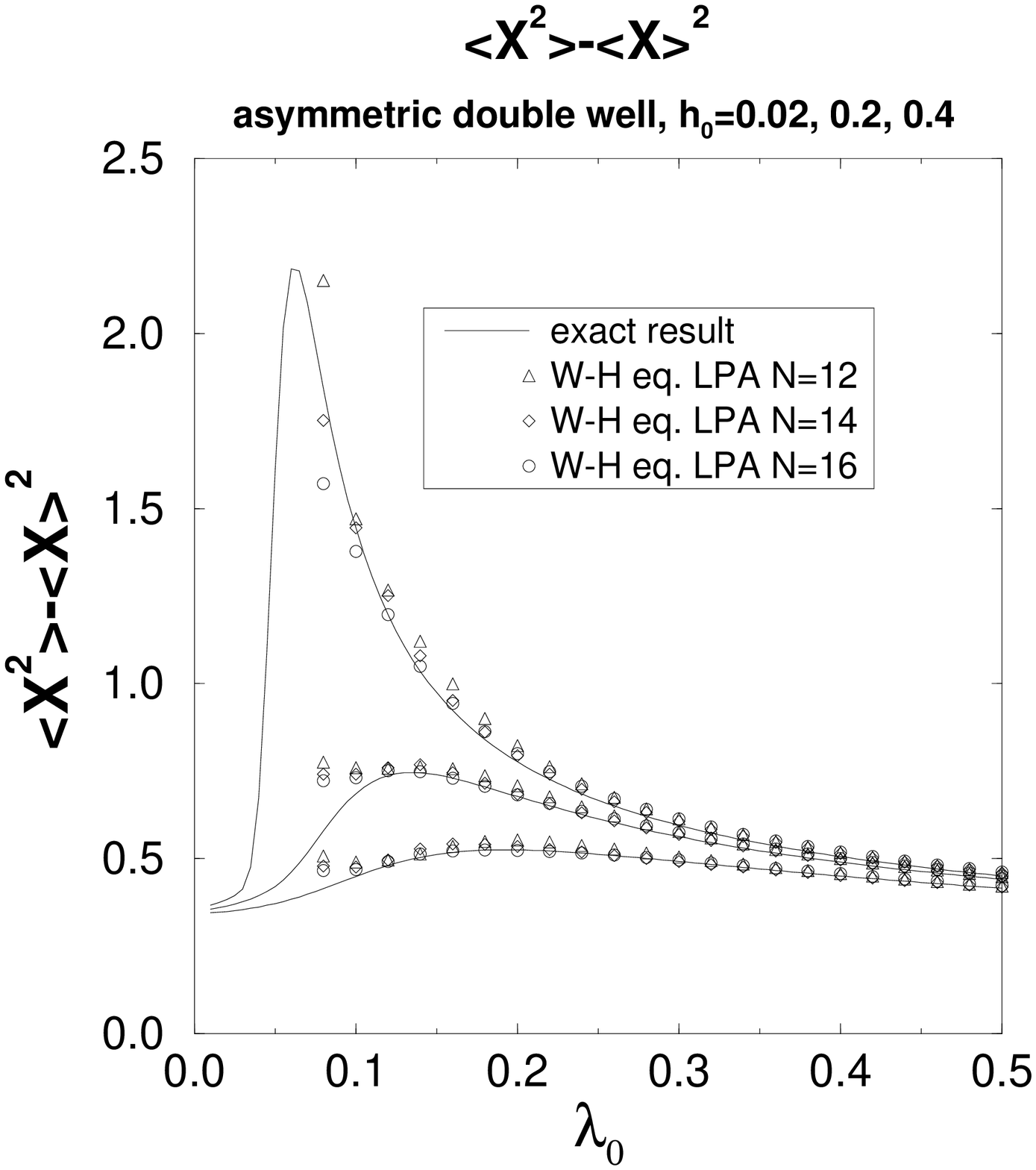}
\vspace{-3mm}
 \caption{$\langle \Omega |\hat{x}^2| \Omega \rangle_{c}$
}
 \label{fig:asymXDN}    
}
\end{figure}

We display results for other quantities 
in Figure.\ref{fig:asymXN} and Figure.\ref{fig:asymXDN}
for three values of $h_0$, 
from top to bottom, $h_0=0.02,\!~0.2,\!~0.4$.
The expectation value of $\hat{x}$,
$M_1 =\langle \Omega |~\hat{x}| \Omega \rangle$,
is shown in Figure.\ref{fig:asymXN},
and the variance of $\hat{x}$ is shown in Figure.\ref{fig:asymXDN}.
The NPRG results appear to be perfect on the strong coupling side,
while they are incorrect in the weak coupling region.
\par
\par
\section{Applications to various quantum systems}
We have seen that the NPRG method is very effective in analyses 
of non-perturbative dynamics in quantum mechanical systems.
Here, we apply the NPRG method to more non-trivial quantum systems.
\subsection{Supersymmetric quantum mechanics}
Here we analyze supersymmetric theory, in which the non-perturbative 
dynamics of the system are crucial.
We consider the 
SUSY QM theory, which was introduced by Witten as a toy model 
for dynamical SUSY breaking.\cite{wi1,sv} 
The Hamiltonian is given by 
\begin{eqnarray}
\hat{H}&=&\frac12\left[\hat{P}^2+\hat{W}^2(x)
+\sigma_3\frac{d\hat{W}(x)}{dx}\right]=
\left(\matrix{\hspace{-10mm}
\frac{1}{2}\hat{P}^2
+\hat{V}_{+}(x)~~~~~~~~~~0
\cr 
~~~~~~~~0~~~~~~~~~~~~
\frac{1}{2}\hat{P}^2+\hat{V}_{-}(x)}\right) ,~~~~~~~~ \\ \label{(38)}
\hat{V}_{\pm}(x)&\equiv&\frac{1}{2}\hat{W}^2(x)\pm
\frac{1}{2}\frac{d\hat{W}(x)}{dx},\label{(38a)}
\end{eqnarray}
where $\hat{W}(x)$ is called the SUSY potential. 
We define the super charges
\begin{eqnarray}
\hat{Q}_1&=&\frac12(\sigma_1\hat{P}+\sigma_2\hat{W}(x)),\\ \label{(38b)}
\hat{Q}_2&=&\frac12(\sigma_2\hat{P}-\sigma_1\hat{W}(x)),\label{(38c)}
\end{eqnarray}
and the Hamiltonian is written as
\begin{eqnarray}
\hat{H}=\{\hat{Q}_1,\hat{Q}_1\}=\{\hat{Q}_2,\hat{Q}_2\}.\label{(38d)}
\end{eqnarray}
This ensures that the vacuum energy is always non-negative: 
\begin{eqnarray}
E_0=\langle\Omega|
\hat{H}|\Omega\rangle
     =2\left\Vert\hat{Q}_1|\Omega\rangle\right\Vert ^{2}
     =2\left\Vert\hat{Q}_2|\Omega\rangle\right\Vert ^{2}\geq0.\label{(38e)}
\end{eqnarray}
\par
The vacuum energy $E_0$ is the order parameter of dynamical SUSY
breaking; that is,
\begin{eqnarray}
E_0=0 \quad \Rightarrow
        &&~~~\hat{Q}_1|\Omega\rangle=0 , ~\hat{Q}_2|\Omega\rangle=0 \qquad
        {\rm unbroken ~~SUSY}, \nonumber\\
E_0>0 \quad \Rightarrow
        &&~~~\hat{Q}_1|\Omega\rangle\ne0 , ~\hat{Q}_2|\Omega\rangle\ne0
        \qquad {\rm broken ~~SUSY}.\nonumber
\end{eqnarray}
Furthermore, the perturbative corrections to $E_0$ are 
vanishing for any order of the perturbation. 
This is known as the non-renormalization theorem. 
In fact, with the SUSY potential $W(x)=gx^2-x$,
the potential $V_{+}(x)$ becomes
\begin{eqnarray}
V_+(x) =~~\frac{1}{2}g^2x^4-gx^3+\frac{1}{2}x^2+gx-\frac{1}{2}.\label{(39)}
\end{eqnarray}
The perturbative corrections to the energy spectrum
are calculated as  
\begin{eqnarray}
E_n=n&+&\frac38g^2[2n^2+2n+1]-\frac38g^2[10n^2+2n+1] \nonumber \\
&-&\frac{1}{32}g^4[34n^3+51n^2+59n+21]
+\cdots .\label{(40)}
\end{eqnarray}
These corrections to $E_0$ are canceled out at each order
of $g$, and thus there are no perturbative corrections. 
Hence, a non-vanishing $E_0$ is realized only through
non-perturbative effects caused by the essential singularity 
at the origin of the coupling constant. 
\par
\begin{figure}[htb]
\hspace{0mm}
 \parbox{65mm}{
 \epsfxsize=65mm     
 \epsfysize=65mm
  \leavevmode
\epsfbox{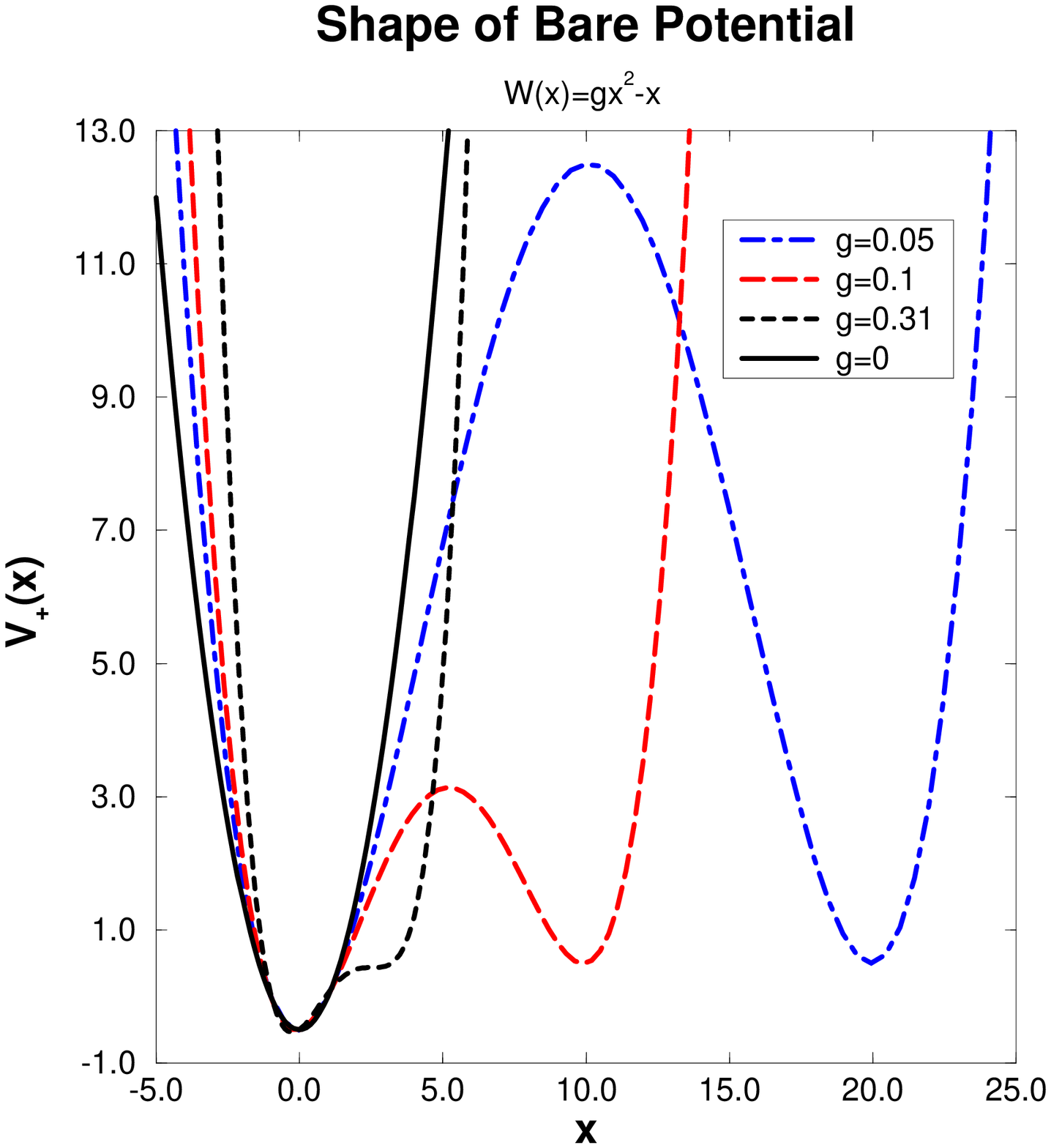}
\vspace{-3mm}
 \caption{Bare potentials.}  
 \label{fig:bpote}        
 }
\hspace{0mm} 
\parbox{65mm}{
 \epsfxsize=65mm     
 \epsfysize=65mm
 \leavevmode
\epsfbox{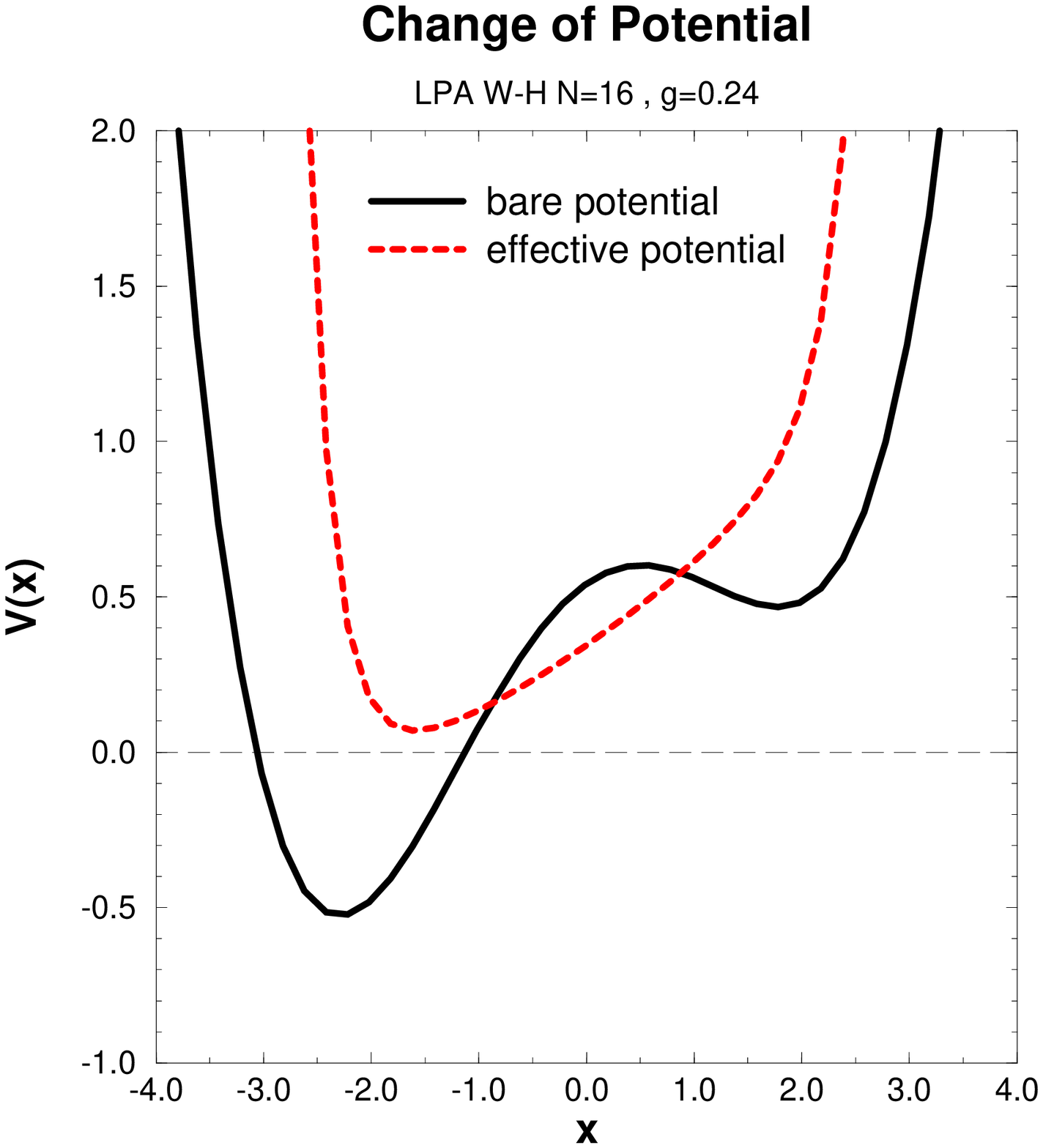}
\vspace{-3mm}
 \caption{Potential flow.}
 \label{fig:epote}       
 }
\end{figure}
\par
We analyze this system using the LPA W-H equation
with $N=16$ for the operator expansion.\cite{ahtt} 
We calculate the effective potential 
for a wide range of values of the parameter $g$. 
The case of vanishing $g$ corresponds to the harmonic oscillator 
with a constant term $-\frac{1}{2}$, and 
SUSY does not break in this case ($E_0=\frac{1}{2}-\frac{1}{2}=0$). 
However, SUSY is dynamically broken 
for any non-vanishing $g$. 
Note that for small $g$, the bare potential is 
an asymmetric double-well, while 
for $g > \sqrt[4]{\frac{1}{108}}\simeq 0.31$, 
it is a single-well, and quantum tunneling is 
irrelevant (Figure.\ref{fig:bpote}). 
For any value of $g$, the minimum of the bare potential $V_{+}$
is at $x=0$. 
Figure.\ref{fig:epote} displays the result for
$g$=0.24, where the effective potential evolves into a convex form, and
its minimum turns out to be positive; 
that is, our NPRG method gives a positive $E_0$ correctly, 
and describes the dynamical SUSY breaking.        
\par
\begin{figure}[htb]     
\hspace{0mm}
 \parbox{65mm}{
 \epsfxsize=65mm     
 \epsfysize=65mm
  \leavevmode
\epsfbox{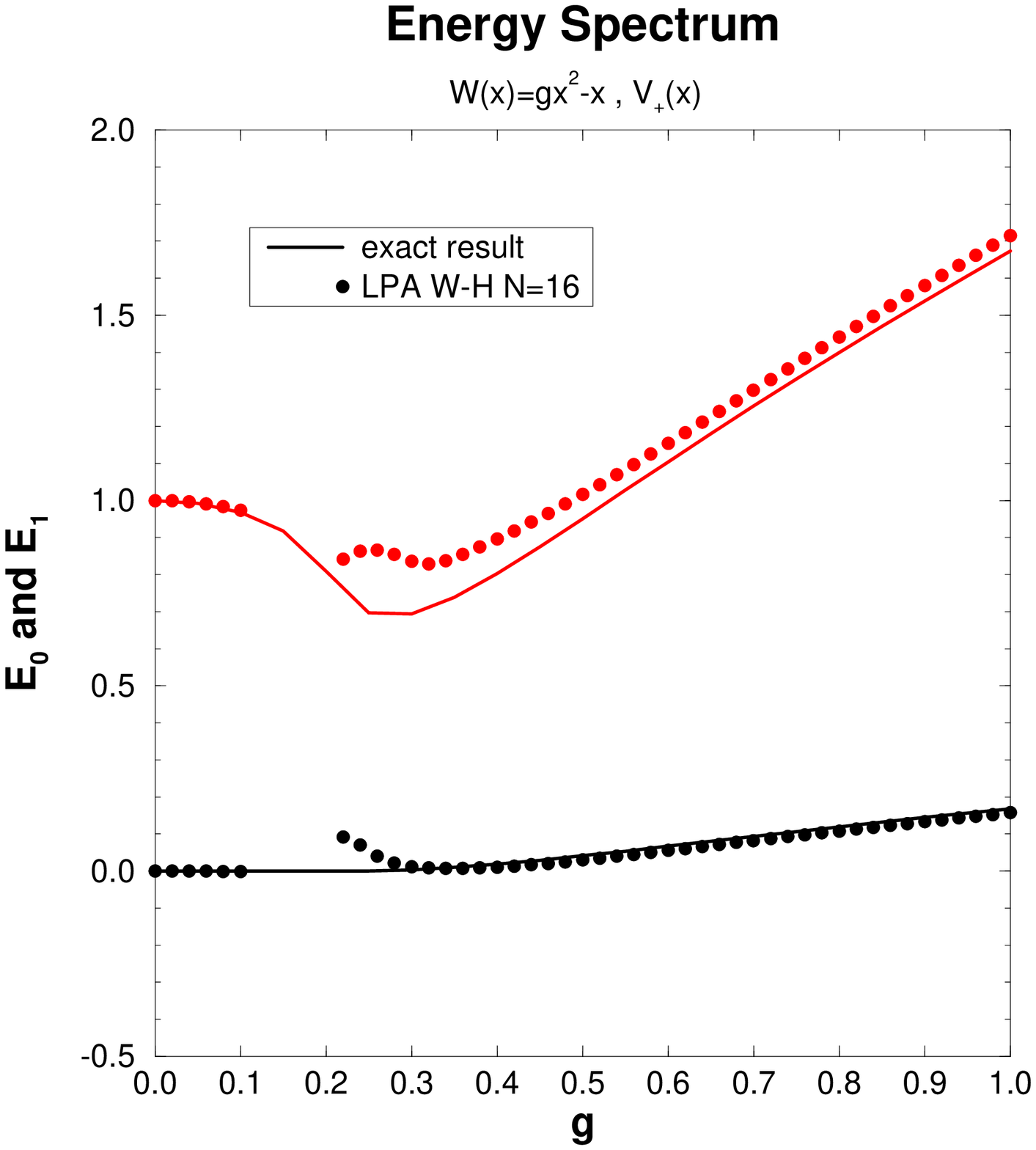}
\vspace{-3mm}
 \caption{Energy spectrum.} 
 \label{fig:rspe}       
 }
\hspace{0mm} 
\parbox{65mm}{
 \epsfxsize=65mm      
 \epsfysize=65mm
 \leavevmode
\epsfbox{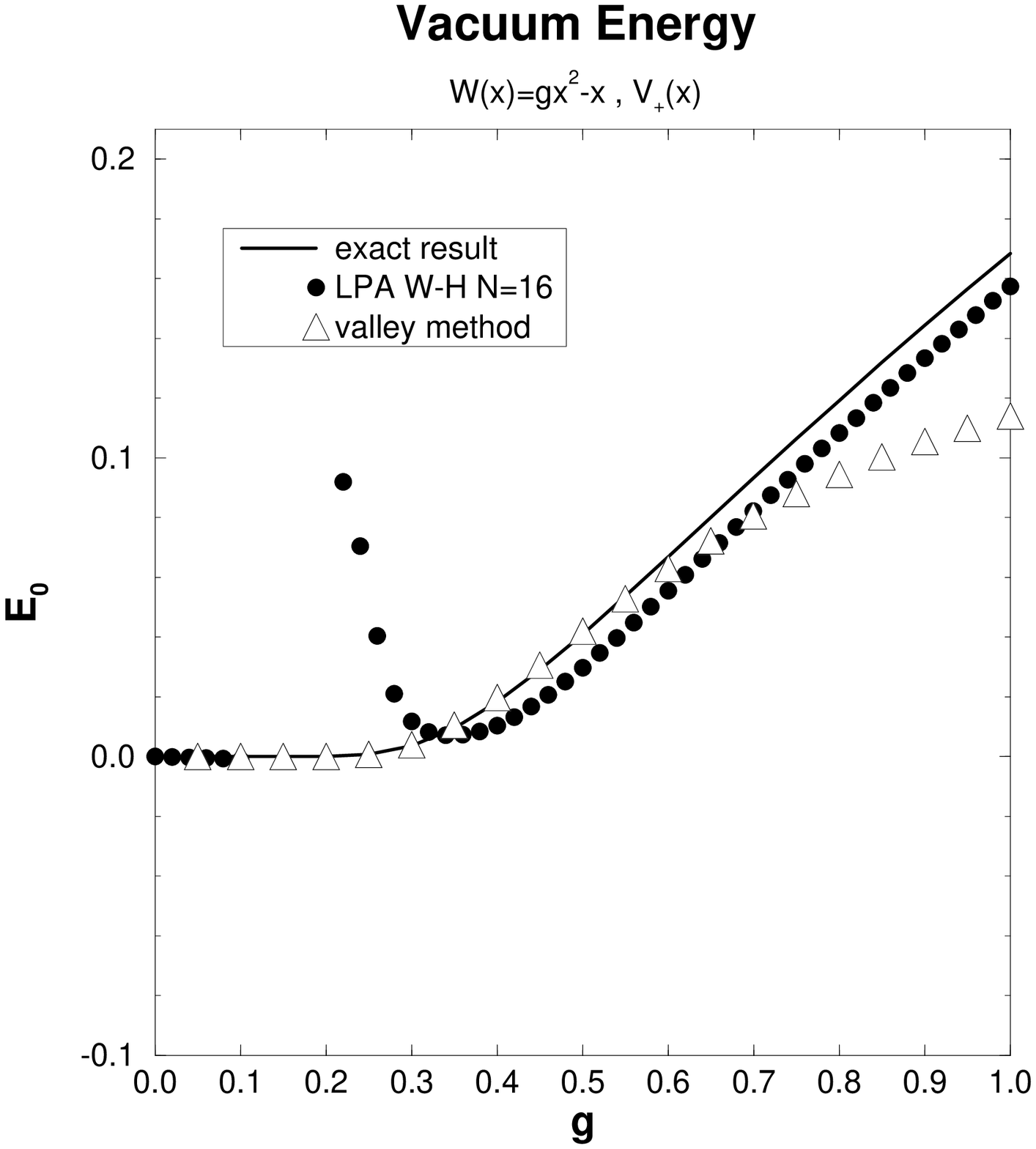}
\vspace{-3mm}
 \caption{NPRG and valley method.}
 \label{fig:vspe2}      
 }
\end{figure}
\par
As is shown in Figure.\ref{fig:rspe}, the NPRG results
are excellent in the weak coupling region and
strong coupling region, but not in the region 
where the bare double-well potential becomes deep.
In this intermediate region ($ 0.1\lsim g\lsim 0.2$), 
we cannot obtain reliable results because of large numerical errors,
while the valley method
works very well, as shown in Figure.\ref{fig:vspe2}. 
The valley method evaluates the ground state energy as  
$E_0=\frac{1}{2\pi}e^{-\frac{1}{3g^2}}$ and reproduces 
the exact value in the weak coupling region.\cite{ao2}
However, it does not work in the strong coupling region 
($g>0.31$), where the valley instanton is no longer 
a good approximate solution of the valley equation. 
Again, we find that the two methods are complementary.
\subsection{Two particle systems}
Next, we apply the NPRG method to quantum many particle systems.
As the simplest system, we analyze two particle ($\varphi_1,\varphi_2$) 
dynamics with the following potential $V_0(\varphi_1,\varphi_2)$:
\begin{equation}
 V_0(\varphi_1,\varphi_2)
=-\frac{1}{2}\varphi^2_1+\lambda_0 \varphi_1^4-\frac{1}{2}
\varphi_2^2+\lambda_0
 \varphi_2^4+F(\varphi_1,\varphi_2). \label{(41)}
\end{equation}
Without the interaction $F(\varphi_1,\varphi_2)$ 
between the two particles, 
the four degenerate ground states  
are mixed by tunneling, splitting into three
$\varphi_1\leftrightarrow \varphi_2$
symmetric states 
and one anti-symmetric state. 
For the interaction $F(\varphi_1,\varphi_2)$, we now choose  
$\varphi_1\leftrightarrow \varphi_2$ symmetric interactions
and investigate how this interaction affects
the energy levels of three symmetric states. 
\par
The LPA W-H equation for two particles is written  
 \begin{equation}
 \Lambda\pa{V_{\Lambda}}{\Lambda}
=-\frac{1}{2\pi}~\!\Lambda~\!
  {\rm Tr}\log\ska{\delta_{ab}+\papa{{V}_{\Lambda
  }}{\varphi_{a}}{\varphi_{b}}}, \label{(42)}
 \end{equation}
where ``Tr'' represents the trace over the subscripts $a,b$ which
correspond to the two particles.
We consider three types of interactions,
 \begin{equation}
F(\varphi_1,\varphi_2)=C\varphi_1\varphi_2~,
~C_2(\varphi_1-\varphi_2)^2~,~C_4(\varphi_1-\varphi_2)^4.\label{(43)}
 \end{equation}
In the cases of the second and third types, for
$C_2, C_4 >0$ the interaction is attractive, and for 
$C_2, C_4 <0$ it is repulsive.
Since we now treat only the ($\varphi_1\leftrightarrow \varphi_2$)
state, it is convenient to convert the variables 
from ($\varphi_1,\varphi_2$) to ($x_1,x_2$) as follows: 
\begin{equation}
\left(\matrix{x_1 \cr x_2}\right)=\frac{1}{\sqrt{2}}\left(\matrix{1-1\cr1~~~~1
}\right)
\left(\matrix{\varphi_1 \cr \varphi_2}\right).\label{(44)}
\end{equation}
The LPA W-H equation for $(x_1,x_2)$ has the same form as (\ref{(42)}). 
The lowest energy splitting for symmetric state, 
$\Delta E=E_1-E_0$, is expressed in terms of the effective mass of $x_2$.
Of course, in the $C=0$ case, this is equal to the effective mass in
one particle system.   
\par
The bare potentials are written, corresponding to (\ref{(43)}), as
 \begin{eqnarray}
 V_0(x)&=&\frac{1}{2}(-1-C)x^2_1+\frac{\lambda_0}{2} x_1^4
+\frac{1}{2}(-1+C)x_2^2+\frac{\lambda_0}{2}
 x_2^4+3\lambda_0 x_1^2x_2^2 ,\label{(45)}\\
 V_0(x)&=&\frac{1}{2}(-1+4C_2)x^2_1+\frac{\lambda_0}{2} x_1^4
-\frac{1}{2}x_2^2+\frac{\lambda_0}{2}
 x_2^4+3\lambda_0 x_1^2x_2^2 ,\label{(46)}\\
 V_0(x)&=&-\frac{1}{2}x^2_1+\left(\frac{\lambda_0}{2}+4C_4\right) x_1^4
-\frac{1}{2}x_2^2+\frac{\lambda_0}{2}
 x_2^4+3\lambda_0 x_1^2x_2^2.\label{(47)}
 \end{eqnarray}
We analyze these systems for small $C,C_{2}$ and $C_{4}$.
We set $\lambda_0=0.2$, which is in the parameter region 
where the NPRG works perfectly in previous analyses. 
The LPA W-H equation was solved numerically 
using the operator expansion with $N=12$.
We also calculated $\Delta E$
from the first order perturbation theory with 
one particle Schr\"odinger wave functions.
\par
\begin{figure}[htb]
\begin{center}
\epsfysize=65mm
\epsfxsize=65mm
\leavevmode
\epsfbox{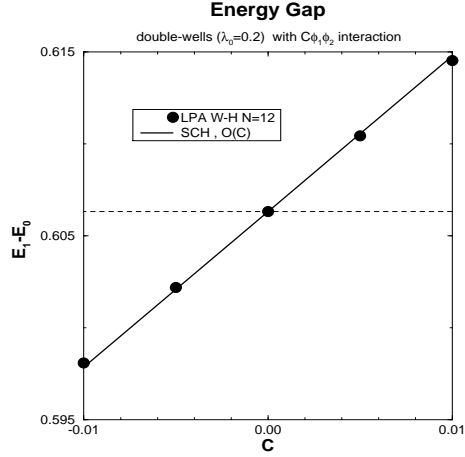}
\vspace{-3mm}
\caption{Linear interaction.}
 \label{fig:cm1}
\end{center}
\end{figure}
\begin{figure}[htb]
\hspace{0mm}
 \parbox{65mm}{
 \epsfxsize=65mm     
 \epsfysize=65mm
  \leavevmode
\epsfbox{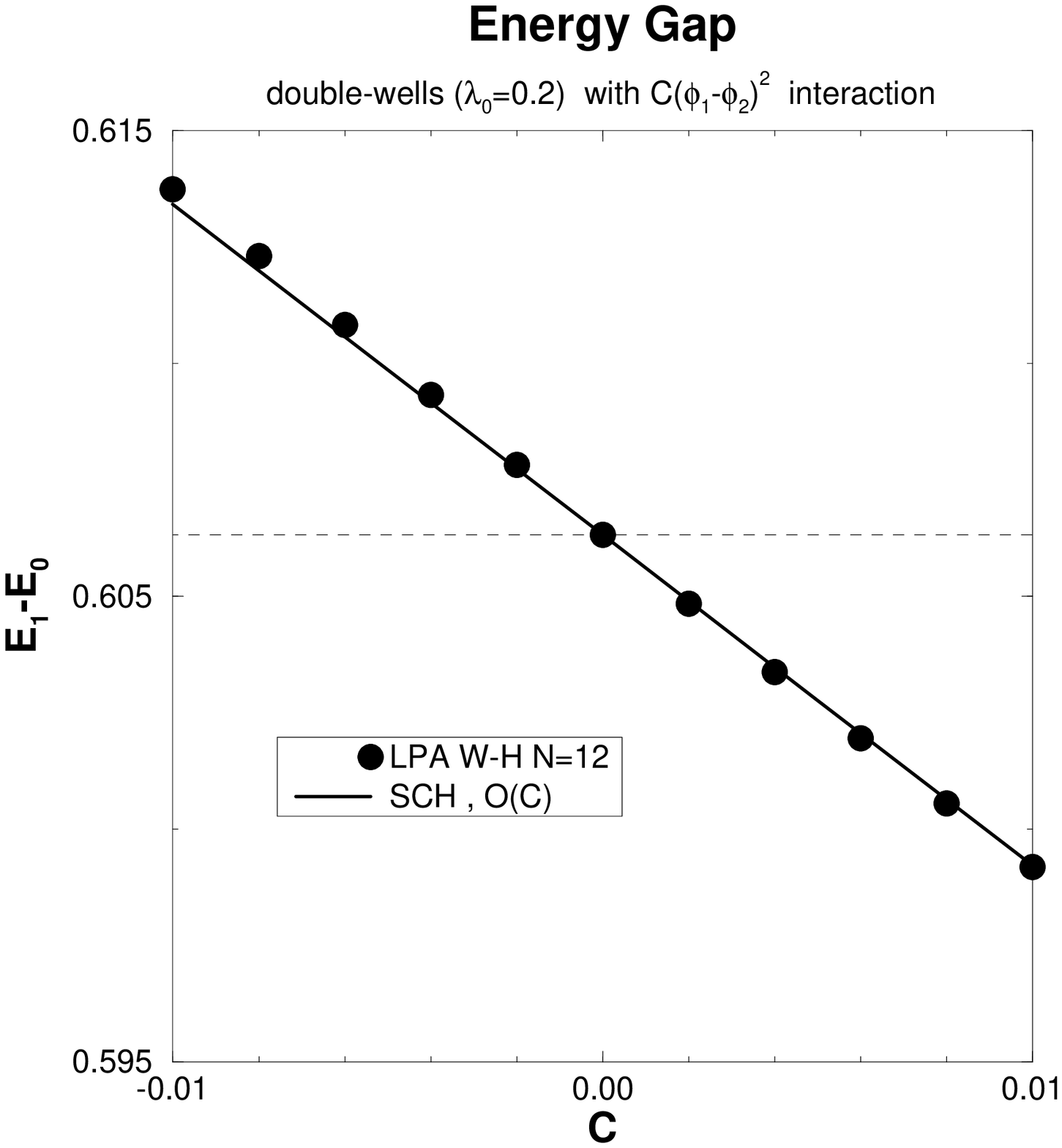}
\vspace{-3mm}
 \caption{Quadratic interaction.}  
 \label{fig:cm2}        
 }
\hspace{0mm} 
\parbox{65mm}{
 \epsfxsize=65mm     
 \epsfysize=65mm
 \leavevmode
\epsfbox{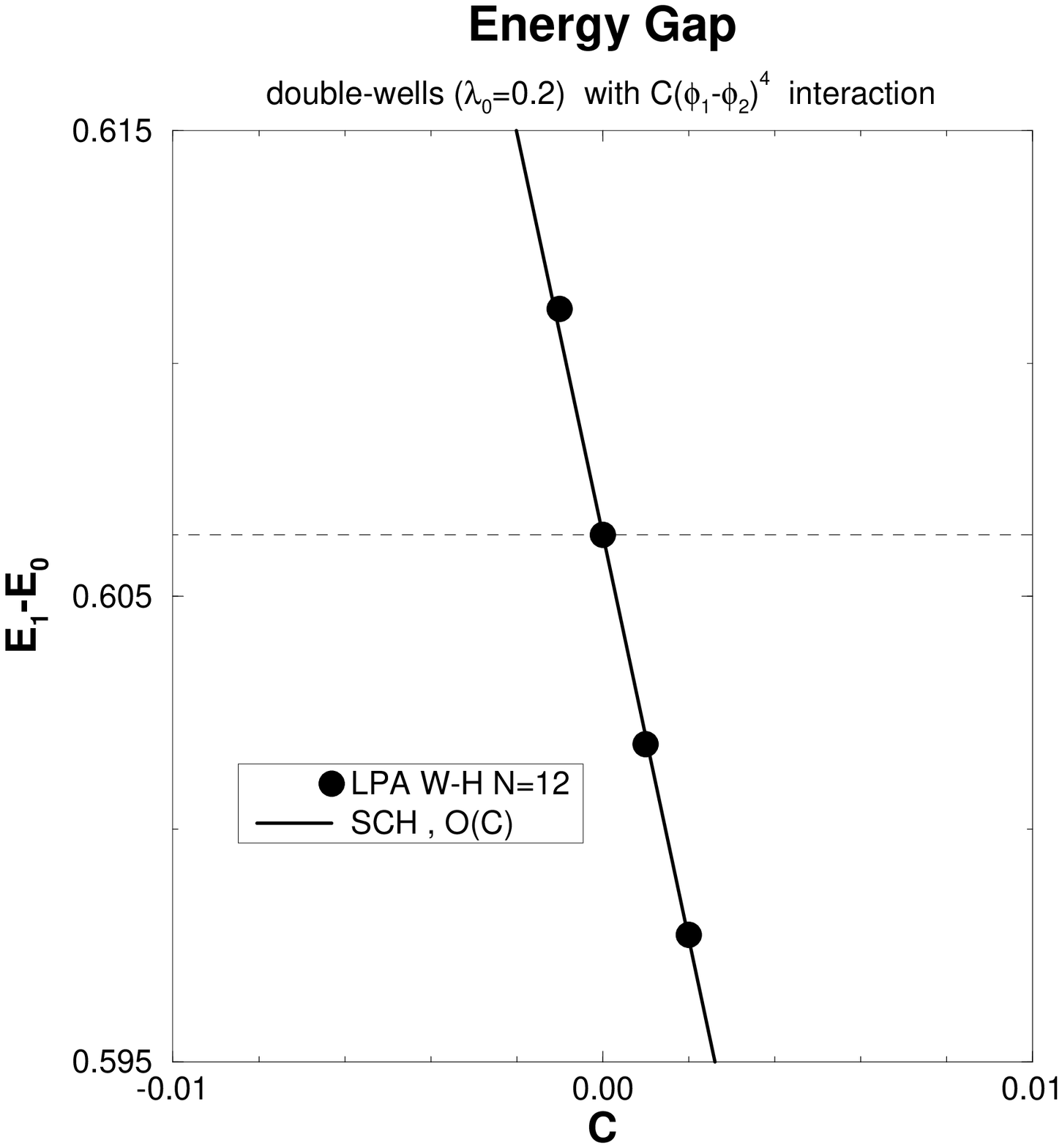}
\vspace{-3mm}
 \caption{Quartic interaction.}
 \label{fig:cm4}       
 }
\end{figure}

The results for small $C,C_{2}$ and $C_{4}$ are shown in
Figures \ref{fig:cm1}, \ref{fig:cm2} and \ref{fig:cm4}.
We see that the NPRG results and 
the Schr\"odinger wave function results are
almost the same in these small interaction regions.  
These results indicate that 
an attractive interaction ($C_2, C_4 >0$) causes $\Delta E$
to decrease, and a repulsive interaction ($C_2, C_4 <0$) causes
it to increase.
\par
\vspace{0mm}
\par
Here we have shown that for multi-particle systems with interactions,
the NPRG method can be applied equally without any change of formulation.
We are now carrying out calculations to obtain 
non-trivial relations between 
particle interactions and tunneling enhancement/suppression.
These results will be reported elsewhere.
\section{Summary and Outlook}
We have applied the NPRG method to various quantum systems 
and used it to analyze non-perturbative physics.
Even in the first stage of approximation, LPA, 
we successfully evaluated the non-perturbative
quantities that should be given by the 
summation of all orders of the diverging perturbative series.
We also found that for non-perturbative quantities 
characterized by an essential singularity, 
the LPA W-H equation again works very well in the region where
the instanton-type method breaks down, i.e. the strong
coupling region. 
However, NPRG is not effective in the weak coupling
region, due to large numerical errors.
In these regions, the approximation used to solve the NPRG equation 
should be improved in order to obtain correct results.
To summarize, the NPRG method and the instanton (or valley) method
play complementary roles. 
Also, from a practical point of view, the NPRG method is
a useful new tool for analysis of various quantum systems
in a wide parameter region. 
We have obtained good non-perturbative results for SUSY QM. 
We also showed that interacting quantum particles can be treated 
in a similar way.
\par
In the flow diagrams, we observed singular behavior in the small coupling
region, and found that it becomes more singular under low-order truncation
of the operator expansion.
The origin of the difficulty 
which we encounter in our NPRG analysis 
resides in the approximation scheme we employed.
We must develop `better' approximations, which may
depend on the individual systems under study. 
We also need to study in detail 
how to extract physical information
from the effective potential and the effective action.   

%\section*{Acknowledgments}
% \pagestyle{myheadings}                      
%We would like to thank ...

\appendix\section{First Pole Dominance}
Here we confirm the first pole dominance in the two-point function 
of anharmonic oscillators.
In the local potential approximation, the two-point function 
is given by following
 \begin{eqnarray}   
\sum_{n=1}^{\infty}\frac{D_n}{E^2+(E_n-E_0)^2}
&\stackrel{LPA}{=}&
\frac{1}{E^2+m^2_{\rm eff}}.\nonumber
 \end{eqnarray}
This substitutes one pole for an infinite number of poles.
Therefore, if the multi-pole contribution becomes significant, 
the correspondence (\ref{(23)}) must be wrong.
\begin{figure}[htb]
\hspace{0mm}
 \parbox{65mm}{
 \epsfxsize=65mm     
 \epsfysize=65mm
  \leavevmode
\epsfbox{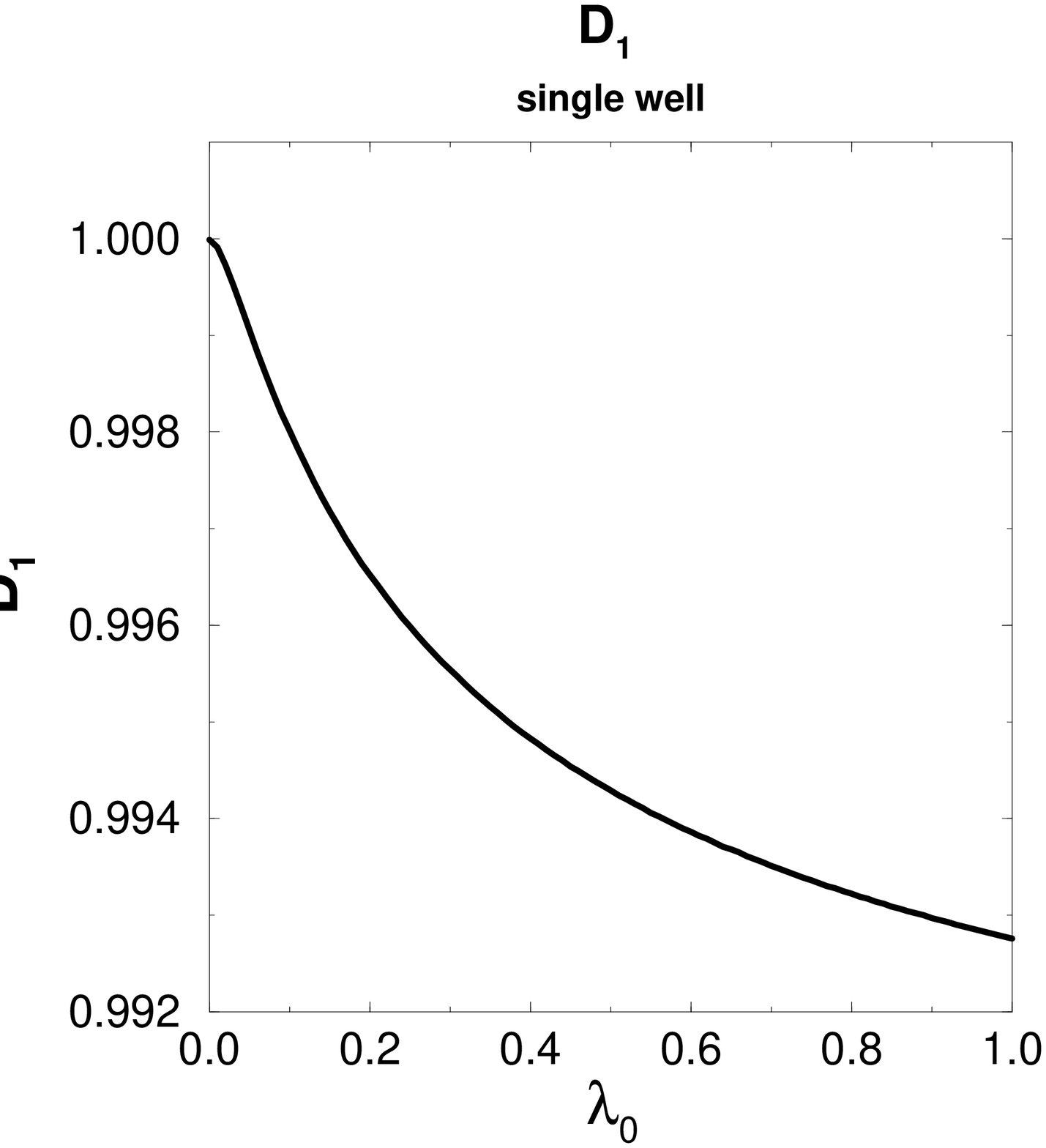}
\vspace{-5mm}
 \caption{$D_1$ for single well potential.}  
 \label{fig:D1s}        
 }
\hspace{0mm} 
\parbox{65mm}{
 \epsfxsize=65mm     
 \epsfysize=65mm
 \leavevmode
\epsfbox{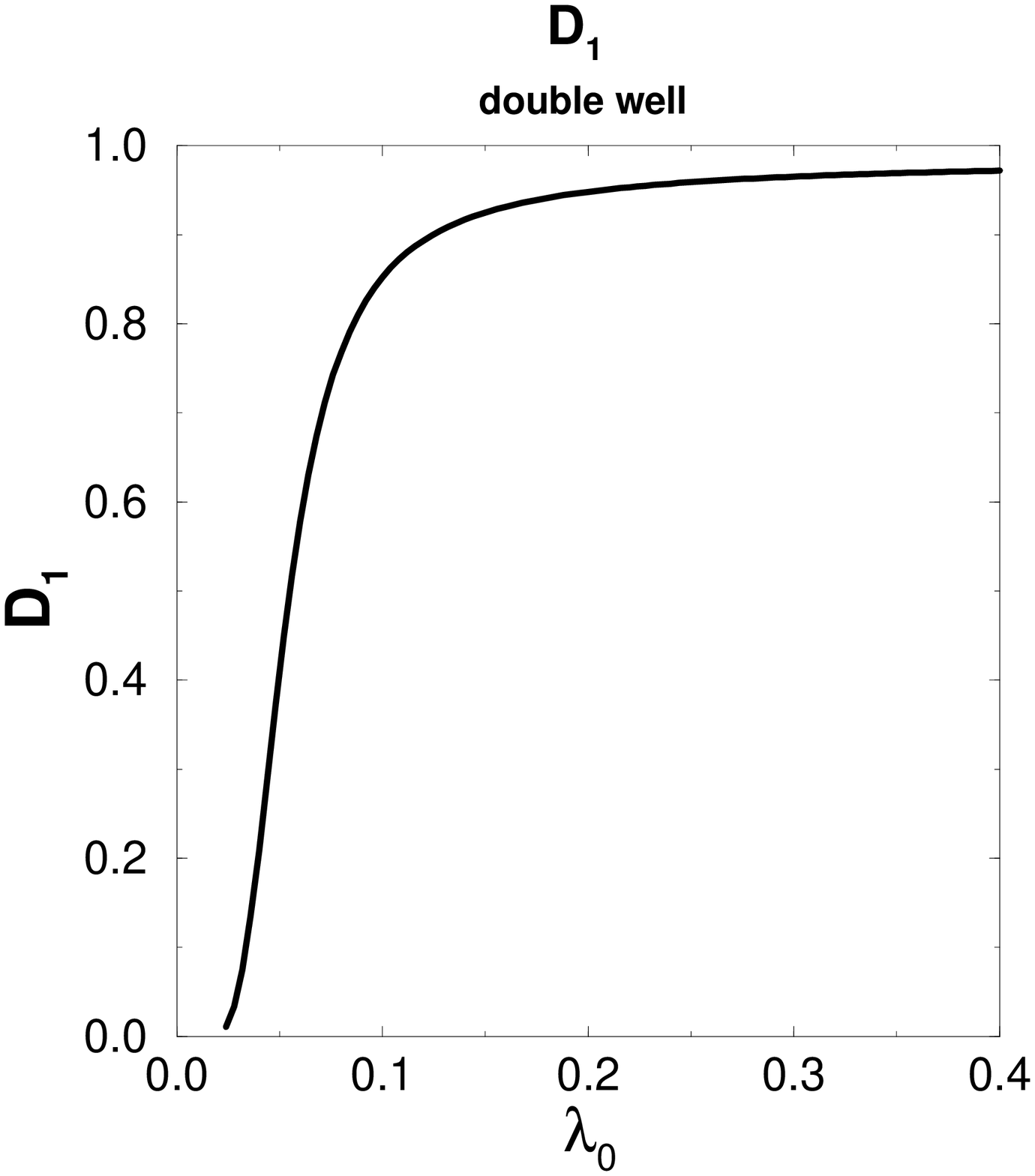}
\vspace{-5mm}
 \caption{$D_1$ for double well potential.}
 \label{fig:D1d}       
 }
\end{figure}
We evaluated the first pole coefficient $D_1$ 
for a single-well (\ref{(30)}) and a double-well (\ref{(32)}) 
by solving the Schr\"odinger equation numerically.
The results are displayed in Figures \ref{fig:D1s} and \ref{fig:D1d}.
\par
We should note that the relation $\sum_n D_n=1$ always holds.
For the single-well potential, the first pole dominates
almost completely.
This corresponds to the fact that the results obtained with the
LPA W-H equation
reproduce the correct results. 
On the other hand, for the double-well potential,
the first pole dominance begins to disappear in  
the region near $\lambda _0 = 0.1 - 0.15 $, where
the results obtained with the LPA W-H equation become poor.
%
%----------------------------------------------------------------------

\end{document}